\documentclass[aps,pra,floatfix,twocolumn,nofootinbib,showpacs]{revtex4}
\usepackage{graphicx} \usepackage{amsmath} \usepackage{amssymb}
\newcommand{\comment}[1]{}
\newcommand{\BEQ}{\begin{equation}}
\newcommand{\EEQ}{\end{equation}}
\newcommand{\BEA}{\begin{eqnarray}}
\newcommand{\EEA}{\end{eqnarray}}

\renewcommand{\S}{{\bf S}}
\newcommand{\R}{{\bf R}}

\newcommand{\bea}{\begin{eqnarray}} 
\newcommand{\eea}{\end{eqnarray}}

\newcommand{\al}{\alpha}
\newcommand{\K}{K}
\newcommand{\I}{\mathbb{I}}
\newcommand{\lb}{\left[}
\newcommand{\rb}{\right]}

\begin{document}

\title{Thermodynamic limits of dynamic cooling}
\author{ Armen E. Allahverdyan$^1$, Karen V. Hovhannisyan$^1$, Dominik Janzing$^2$, Guenter Mahler$^3$}
\affiliation{$^1$Yerevan Physics Institute, Alikhanian Brothers Street 2, Yerevan 375036, Armenia,\\
$^2$MPI for Intelligent Systems, Spemannstrasse 38 72076 Tuebingen, Germany,\\
$^3$Institute of Theoretical Physics I, University of Stuttgart, Pfaffenwaldring 57, 70550 Stuttgart, Germany}

\begin{abstract} We study dynamic cooling, where an externally driven
two-level system is cooled via reservoir, a quantum system with initial
canonical equilibrium state.  We obtain explicitly the minimal possible
temperature $T_{\rm min}>0$ reachable for the two-level system. The
minimization goes over all unitary dynamic processes operating on the
system and reservoir, and over the reservoir energy spectrum. The
minimal work needed to reach $T_{\rm min}$ grows as $1/T_{\rm min}$.
This work cost can be significantly reduced, though, if one is satisfied
by temperatures slightly above $T_{\rm min}$. Our results on $T_{\rm
min}>0$ prove unattainability of the absolute zero temperature without
ambiguities that surround its derivation from the entropic version of
the third law. The unattainability can be recovered, albeit via a
different mechanism, for cooling by a reservoir with an initially
microcanonic state. We also study cooling via a reservoir consisting of
$N\gg 1$ identical spins. Here we show that $T_{\rm
min}\propto\frac{1}{N}$ and find the maximal cooling compatible
with the minimal work determined by the free energy.

\end{abstract}








\maketitle

\section{Introduction}

Some physical systems have to be cooled before they can demonstrate
interesting features, e.g. quantum properties of matter are typically
displayed only after suitable cooling. There are various cooling
methods; we distinguish here the "brute force method", the Nernst-set
up, and the so-called dynamic cooling.  The brute force method of
cooling amounts to bringing the system in contact with a low-temperature
thermal bath, so that it relaxes to this lower temperature. 

When trying to do without a pre-existing cold bath different strategies are needed:
Low temperatures can alternatively be produced dynamically from an
initially equilibrium system via cyclic action of an external field
\cite{callen,wheeler,landsberg,grif,klein,callen_note,falk,magnetocaloric,abo,ernst,ole,slichter,suter,exp,algol,schulman,a,lasercooling}.
In this case the resulting cooled state should be exploited before the
system has time to relax back to equilibrium.

Within macroscopic quasi-equilibrium thermodynamics some of these
methods have been summarized in the Nernst set-up of cooling, which
supports a formulation of the third law
\cite{callen,wheeler,landsberg,grif,klein,callen_note,falk}.  In fact,
this law controls the cooling of a macroscopic target initially in
contact with an appropriate macroscopic reservoir. The target is then
typically subject to a two-step cooling process, in which an external
field cycle is executed under (quasi-equilibrium) isothermal and
adiabatic conditions, respectively.  A well-known realization of this
Nernst set-up is the magnetocaloric effect which was first observed in
1880, but still attracts attention \cite{magnetocaloric}. 

There are however other methods of cooling that became important with
the rise of low-temperature physics
\cite{abo,ernst,ole,slichter,exp,algol,schulman,a,lasercooling}. Here
the target of cooling (or the reservoir, or both) is generally not a
macrosocpic system, while the process that produces low temperatures ceases to be a
quasi-equilibrium one. Hence the understanding of these methods should
rely on the actual dynamics rather than on quasi-equilibrium
thermodynamics. We therefore refer to them as dynamic cooling. A notable
example of this is the dynamical nuclear polarization in NMR
\cite{abo,ernst,ole,slichter,exp}. Within this method the nuclear spins
are cooled via transferring polarization from electron spins by means of
external microwave fields \cite{abo,ernst,ole,slichter,exp}. It was
originally employed in the solid-state NMR, but since recently it is
also applied for the liquid-state NMR in view of its medical and
bio-physical applications \cite{exp}. Other examples of dynamic cooling
are algorithmic cooling \cite{algol,schulman}, bath-assisted cooling
\cite{a} and laser cooling of motional state in atoms
\cite{lasercooling}. 

Dynamic nuclear polarization illustrates the basic ingredients of
dynamic cooling processes that are seen already in the Nernst set-up:
the target system to be cooled (nuclear spins), the reservoir, which
plays the role of a polarization source or the entropy sink (electrons),
and external fields that couple these two together (microwave radiation
at suitable frequency). Recall that no
cooling is possible without reservoir \cite{nocooling}; i.e.,
one can never cool the whole system (= target + reservoir) by coherent
fields.  
\comment{ For algorithmic cooling the reservoir is a system of nuclear
spins, whose initial polarization is transferred to the target spin
\cite{algol,schulman}.  For bath-assisted cooling the reservoir is a
phononic thermal bath \cite{a}, while for the laser cooling the
ingredients are, respectively, motional states of atoms, vacuum modes
that accept spontaneously emitted photons and laser fields that drive
atoms \cite{lasercooling}. }

As already indicated, the operation of cooling and the concept of the 
third law are intimately interrelated. While we do not attempt to dwell 
on the general ideas behind operationalism as originally promoted by 
Bridgman \cite{bri}, it is nevertheless worthwhile to stress the 
potential benefits of such an approach even to modern thermodynamics, and, 
in particular, for reaching the objectives of this paper. Our aim here is 
threefold: To formalize cooling schemes in terms of imposed limited 
resources, to parametrize specific reservoir models, and to investigate the 
resulting minimal temperature as a (scaling) function of those operational
parameters. 

In particular, we intend to study the minimal temperature $T_{\rm min}$
reachable within a sufficiently general set-up of dynamic cooling, where
both the target and reservoir are finite quantum systems, and where the
final state of the reservoir can generally be far from equilibrium. We
determine how $T_{\rm min}$ depends on resources of the set-up and note
that the minimal work necessary to achieve $T_{\rm min}$ grows as
$\frac{1}{T_{\rm min}}$ whenever $T_{\rm min}\to 0$. Hence the work is a
diverging resource of dynamic cooling, in contrast to the Nernst set-up
and the third law, where the work done for cooling is always
well-bounded. We also determine the minimal temperature (and the minimal
work needed to attain for it) for several concrete reservoir models. 

Our results on $T_{\rm min}>0$ extend (to dynamic cooling) the
unattainability formulation of the third law.  Hence section \ref{opera}
reviews the (frequently disguised) assumptions of the Nernst set-up and
of the third law in its two formulations. (The reader who is well aware
of the third law may just consult section \ref{summa_contra} for
relevant assumptions.)

The dynamic cooling set-up is defined in section \ref{setup}.  Section
\ref{nozero} shows that dynamic cooling with a reservoir in an initial
canonical state does not allow to reach the absolute zero of temperature
[unattainability]. In section \ref{comparo} we summarize relevant
conditions of the dynamic cooling set-up. 

The following three sections are concerned with reservoir models.
Section \ref{opto} studies the maximal cooling possible within the
dynamic set-up. It confirms that the work is a relevant resource for
cooling and determines the minimal work necessary to achieve the maximal
cooling. Section \ref{homogeneous} studies a reservoir with the
homogeneous, non-degenerate spectrum, while in section \ref{N} the
reservoir is modeled as a thermal bath consisting of $N\gg 1$ spins.
This model is complemented by a discussion on the dynamic cooling
process for initially microcanonic state of the reservoir. 

In the last section we summarize our results and discuss their relations
with other microscopic approaches for studying thermodynamic limits of
cooling: the approach of Ref.~\cite{Dom_Beth}, where the cooling process
is restricted to operations within degenerate subspaces of the joint
Hamiltonian of the system and non-equilibrium reservoir, and the
approach based on quantum refrigerators
\cite{yan,gordon,rezek,segal,karen}. We close by pointing out some open
issues. 

\section{Operational analysis of the Nernst set-up and the third law}
\label{opera}

\subsection{Description of the Nernst set-up}

We are given a macroscopic system in contact with a much larger thermal
bath \cite{callen}. Both have initial temperature $T_{\rm in}$.
One now switches on an external field $g$ acting on the system and
changes it {\it slowly} from its initial value $g_{\rm in}$ to its final
value $g_{\rm fin}$. The contact with the bath is held fixed. Hence this is
an isothermal process. Let $g_{\rm in}$ and $g_{\rm fin}$ be chosen such that
the entropy $S[T,g]$ of the system {\it decreases} (see Fig.~\ref{f00})
\BEA
\label{a1}
S[T_{\rm in}, g_{\rm in}]> S[T_{\rm in}, g_{\rm fin}].
\EEA
The entropy difference is transferred to the bath.

Next, the system is {\it thermally} isolated (i.e., isolated from the
bath) and the field $g$ is {\it slowly} returned back to its original
value $g_{\rm in}$ thereby completing the cooling cycle. (Note: the cyclic condition only refers to the external field, 
i.e. the time-dependent Hamiltonian; it would be meaningless to require the target to return to its initial state.)
Hence this part
of the process is thermally isolated and reversible (adiabatic).  Since
the system is macroscopic, the adiabatic part is characterized by
constant thermodynamic entropy \cite{callen}:
\BEA
\label{a2}
S[T_{\rm fin}, g_{\rm in}]= S[T_{\rm in}, g_{\rm fin}],
\EEA
where $T_{\rm fin}$ is the final temperature of the system.  Due to
(\ref{a1}) and $\frac{\partial S}{\partial T}\geq 0$, we get $T_{\rm
fin}<T_{\rm in}$: some cooling has been achieved
\cite{callen,wheeler,landsberg,grif,klein,callen_note,falk,magnetocaloric}.
Recall that $\frac{\partial S}{\partial T}\geq 0$ is one of basic
conditions of local thermodynamic stability (positivity of the specific
heat) \cite{callen}. The system will stay cold as long as it is well isolated from the bath.

This cooling cycle has a work-cost. In terms of the free energy
\BEA
\label{buka}
F[T,g]=U[T,g]-TS[T,g] 
\EEA
of the system, where $U[T,g]$ is its energy, the work
done in the isothermal step is $F[T_{\rm in}, g_{\rm
fin}]-F[T_{\rm in}, g_{\rm in}]$. The work in the adiabatic step is
$U[T_{\rm fin}, g_{\rm in}]- U[T_{\rm in}, g_{\rm fin}]$. Summing them up
and using (\ref{a2}) one has for the work $W$ invested per cooling cycle
\footnote{We use the sign convention $W>0$, if the work-source looses energy. }:
\BEA
\label{a3}
W=
F[T_{\rm fin}, g_{\rm in}]- F[T_{\rm in}, g_{\rm in}]>0,
\EEA
which is always positive due to $\frac{\partial F}{\partial T}=-S\leq 0$
\footnote{We assume the convention, where the entropy is defined to be
non-negative.} and $T_{\rm in}> T_{\rm fin}$.  Below, in sections
\ref{setup} and \ref{N}, we relate (\ref{a3}) to the minimal work-cost
of the dynamic cooling set-up. Note that $W$ according to (\ref{a3}) is always finite,
even for $T_{\rm fin}=0$. 

\subsection{Entropic formulations of the third law}
\label{3law}

The entropic formulation states that for each fixed and finite value of
the field $g$, the entropy $S[T,g]\geq 0$ {\it smoothly} goes to zero with temperature $T$:
$S[T,g]\to 0$ for $T\to 0$
\cite{callen,wheeler,landsberg,grif,klein,callen_note,falk}; see Fig.~\ref{f00}.   
The entropic formulation is
frequently deduced from the non-degeneracy of the system's ground state.
However, this does not suffice for a derivation, because for a
macroscopic system the appropriate order for taking limits is that first
the thermodynamic limit is taken, the entropy density is calculated, and
then after that the low temperature limit goes \cite{grif,klein}.  Then
the zero-temperature entropy starts to depend on global features of the
energy spectrum.  The full derivation within statistical mechanics 
was attempted several times \cite{grif,klein,callen_note,falk}, but is
regarded to be an open problem, because there are different classes of
theoretical Hamiltonian systems violating the entropic formulation: the
zero-temperature entropy for them is not zero and depends on external
fields \cite{callen_note,grif,casimir1,casimir3,wald,wu,behn,1d}. 

It is believed that Hamiltonians which would violate the entropic
formulation are unstable with respect to small perturbations
\cite{wheeler,callen_note}. Such perturbations would have to be
generated by some appropriate noise model. Unfortunately, a general
recipe for selecting such a model does not seem to be known (or even
exist) \footnote{The reader may consult controversies with the validity
of the entropic formulation in the field of the Casimir effect
\cite{casimir1,casimir2,casimir3} or for black holes \cite{wald},
criteria of this formulation for spin systems \cite{leff,wu}, and
classes of one-dimensional counter-examples to the formulation
\cite{behn,1d}. The authors of \cite{1d} ensured robustness of their
counter-examples to a class of noise models.}. There are results showing
how one should select the noise model for particular cases only
\cite{wheeler,callen_note,casimir1,casimir2,leff,wu}.

\subsection{Unattainability formulation of the third law}

\begin{figure}
\includegraphics[width=7.5cm]{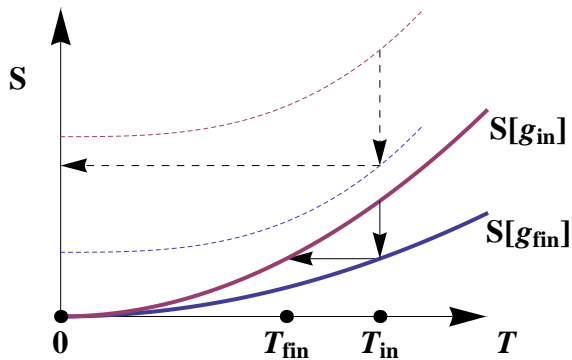}
\vspace{0.65cm}
\caption{ A schematic representation of the relation between the entropic formulation of the third law
and the attainability of $T=0$. $S$ and $T$ denote entropy and temperature, respectively; $g$ is the external field.
Bold (dashed)
lines represent a situation, where the entropic formulation holds (is violated). Arrows represent the isothermal and adiabatic parts
of the cooling cycle. We see that the final temperature $T_{\rm fin}$ is greater than zero. 
However, it is formally possible to attain $T=0$ (dashed arrows) when the entropic formulation 
is violated in the sense that $S[T=0]$ still depends on the field $g$.}
\label{f00}
\end{figure}

Already the entropic formulation of the third law can be brought in
contact with operational requirements. This is even more so for the
unattainability formulation. This formulation states that it is
impossible to reach absolute zero of temperature, $T=0$, by finite
means, i.e. via a {\it tractable} physical process. Note that the
unattainability formulation is clearly different from the entropic
formulation that refers to a limiting feature of a definite function of
temperature (entropy). The unattainability formulation relies on the
notion of tractability, whose {\it general} formalization appears to be
hard to come by. Hence the two formulations cannot be completely
equivalent \cite{landsberg}. 

\comment{Unattainability is {\it not} a well-defined physical concept. It can be
approached in two steps: {\it i)} A parametrization of the underlying
process based on a specific model. {\it ii)} A specification of the
"accessible" parameter window. (Technical constraints do exist, but
might become obsolete in the future. However, there is a tentative
agreement that infinite parameter values like infinite energy input,
infinite system size, infinite precision etc. should never become
accessible.)}

Within the Nernst set-up one naturally assumes that for a tractable
physical process the strength $g$ of the external field should not
assume infinite values \cite{callen,wheeler}. This suffices for a
heuristic derivation of the unattainability formulation from the
entropic formulation; see Fig.~\ref{f00} and \cite{callen}. Indeed,
Fig.~\ref{f00} shows that also shows that the unattainability is
violated together with counter-examples of the entropic formulation
\footnote{
This would not imply that one can ever verify reaching $T=0$, since any
temperature measurement has a finite precision. (The situation with
reaching a positive temperature, say 10 K, is different, because one can
arrange for passing through this temperature at some [uncertain] time.)
In particular, temperature fluctuations can prevent a precise
determination of low temperatures \cite{widom,ja}. According to the
standard thermodynamic approach \cite{widom} these fluctuations will
grow for $T\to 0$ due to the vanishing heat capacity. The issues of
temperature fluctuations was recently reconsidered and clarified in
\cite{ja} from the first principles of quantum mechanics.}. But there
are also examples that would violate the unattainability formulation
only \cite{wheeler,derrida} \footnote{The mechanism of these
counter-examples is that the entropy nullifies at some $T_c(g)>0$ and
stays zero for $T\leq T_c(g)$. }.  Again, one way out of this ambiguous
situation is to look for via "sufficiently" stable Hamiltonians
\cite{wheeler}. 

\comment{
We close this discussion by recalling that the entropic formulation is
sufficient, but not necessary for the unattainability formulation. For
an example consider the Boltzmann gas, whose entropy $S[T,g]$ does not
have a finite limit at $T\to 0$, $S[T\to 0,g]\propto \ln T\to-\infty$,
but $T=0$ cannot be attained, because $S[T\to 0]$ ceases to depend on
(finite) external field $g$ \cite{wheeler}. }

\subsection{Summary of conditions}
\label{summa_contra}

We finally try to summarize the Nernst set-up in terms of two lists of
conditions: The first list {\bf N1-- N6} specifies details of the task: it
reflects our choice for implementing a satisfactory cooling process. In
part these conditions may be tied to the actual technology. The second
list {\bf T1-- T3} introduces fundamental tractability conditions. These
stress the unavoidable finitness of the resources at our disposal and
should remain intact within any future technology. The respective
process will be called tractable insofar as these requirements can be
fulfilled. 

{\bf N1.} Initial state: equilibrium for the system and bath.

{\bf N2.} System (target of cooling): macroscopic.

{\bf N3.} Cooling process: Quasi-equilibrium. 

{\bf N4.} Bath: macroscopic, much larger than the system.

{\bf N5.} External field: changes cyclically. 

{\bf N6.} The Hamiltonian of the target: stable with respect to
perturbations. 

{\bf N1} recalls that we intend to start from a given finite
temperature, while {\bf N2--N4} are assumed because the process is to be
embedded into macroscopic equilibrium thermodynamics. {\bf N2} and {\bf
N3} reflect the technology of Nernst's times. {\bf N5} is assumed,
because the cooled system has to be {\it autonomous}, i.e. once it was
cooled, it can be used in other places, without being kept under a
constant external field.  {\bf N6} is included here because a general
noise model for perturbations is not known: the concrete stability
requirement thus becomes a matter of choice. 

We now turn to the tractability conditions; this list is not necessarily complete (cf. section \ref{micro}).

{\bf T1.} External field has finite strength.

{\bf T2.} The work-cost necessary for cooling is finite.

{\bf T3.} Duration of the cooling cycle is finite.

{\bf T1} turns out to be essential for deducing the unattainability
formulation of the third law. It is also relevant for the entropic
formulation. (It is easy to find examples, where the entropy does not
nullify with the temperature if simultaneously the field goes to
infinity \cite{wheeler}.) Due to (\ref{a3}), {\bf T2} is guaranteed by
{\bf N1--N4}.  Typically, {\bf T3} becomes essential for certain
scenarios: Indeed, in frustrated systems reaching the true equilibrium
may demand unrealistically long observation times; on practically
relevant observation times the system may find itself trapped in
quasi-stationary states, where the non-equilibrium entropy does not go
to zero with the ambient temperature, though the true equilibrium state
does satisfy the entropic formulation. This effect was observed
experimentally \cite{nisoli}. It is also well known for glassy systems
\cite{glass} (residual entropy); see \cite{glass1} for a review. 

\comment{\footnote{
Such effects of incomplete equilibration were anticipated already by
Simon, who excluded them from the formulation of the third law
\cite{wheeler}. This exclusion seems to be however rather formal,
because the above quasi-stationary state may still have all the
operational features of equilibrium, e.g., temperature, stationarity of
observables on experimentally relevant times, {\it etc.}
\cite{glass,glass2}. }.}

\section{Setup for dynamic cooling}
\label{setup}

\subsection{System and reservoir}

A reasonably general set-up for dynamic cooling requires to specify the
class of Hamiltonians for the target system {\bf S} and the reservoir
{\bf R}, together with their initial state. Here, both subsystems will
be taken to be finite-level quantum systems with the overall Hamiltonian
$H_{\rm \bf S+R}$. The initial state of {\bf S+R} is a {\it canonical}
equilibrium one with density matrix
\BEA
\label{mu}
\rho_{\rm \bf S+R}= \frac{e^{-\beta H_{\rm \bf S+R}}}{{\rm tr\,}[\, e^{-\beta H_{\rm \bf S+R}}\,] },
\EEA
and the initial temperature $T=1/\beta$ ($k_{\rm B}=1$). We shall additionally assume that 
in the initial state the coupling between {\bf S} and {\bf R} is negligible:
\BEA
\label{brams}
H_{\rm \bf S+R}=H_{\rm \bf S}+H_{\rm \bf R}, \qquad \rho_{\rm \bf S+R}=\rho_{\rm \bf S}\otimes\rho_{\rm \bf R}.
\EEA
The action of external fields is described by an interaction term in the time-dependent Hamiltonian of {\bf S+R}:
\BEA
\label{ga}
H_{\rm \bf S+R}(t)=H_{\rm \bf S}+H_{\rm \bf R}+V(t),
\EEA
which is switched on at time $t=0$ and switched off at time $t=\tau$: $V(t)=0$ for $t<0$ and for $t>\tau$ ({\it cyclic} motion
of the external field).
$V(t)$ acts on both ${\bf S}$ and ${\bf R}$; if it acts only on ${\bf S}$, no cooling is possible \cite{nocooling}.
$H_{\rm \bf S+R}(t)$ generates a unitary operator $U$
that takes the initial state (\ref{mu}) to the final state of {\bf S+R}:
\begin{gather}
\label{fio}
\rho_{\rm \bf S+R}(\tau)= U\rho_{\rm \bf S+R} U^\dagger,\\
\rho_{\rm \bf S}(\tau)={\rm tr}_{\bf R}\rho_{\rm \bf S+R}(\tau), \qquad 
\rho_{\rm \bf R}(\tau)={\rm tr}_{\bf S}\rho_{\rm \bf S+R}(\tau),
\end{gather}
where we defined also marginal final states of {\bf S} and {\bf R}.

\subsection{Definition of cooling}

How to define cooling of {\bf S} in the non-equilibrium setting? Note
that the maximal cooling is always well-defined, because it means that
the final ground state probability of {\bf S} is [$|0\rangle$ is the ground state of $H_{\bf S}$]
\BEA
p_0(\tau)\equiv \langle 0|\rho_{\rm \bf S}(\tau)|0\rangle =1 \qquad ({\rm or} ~~ T=0). 
\EEA
Defining a
non-maximal cooling via $p_0(\tau)$ is reasonable for two-level systems,
since a larger $p_0(\tau)$ means that the energy distribution in the
final state is more shifted towards the ground state \footnote{One may
define cooling via the maximal eigenvalue of the density matrix,
and not the ground-state probability. In the optimal regime of our
set-up both definitions agree with each other, because the final state 
of {\bf S} is energy-diagonal; see section \ref{maxmin}.}. If in addition the
state of {\bf S} is diagonal in the energy representation, one can
equivalently express the cooling via temperature defined as in
(\ref{mu}). For a multi-level system {\bf S} one needs to take
care in defining the meaning of a possibly non-equilibrium
state of {\bf S} being colder than a given equilibrium state
\footnote{When comparing two systems with the same energy levels|e.g.,
the target of cooling before and after the cooling process realized via
a cyclically changing Hamiltonian|one can define cooling by requiring
that the whole energy distribution is shifted towards the ground state.
In effect, this amounts to using {\it majorization} as measure of
cooling; see \cite{olkin} and Appendix A for the definition of this
concept, see also \cite{Dom_Beth} for a related approach. The drawback
(or viewing differently an advantage) of this definition is that when
the number of energy levels is larger than two, not every two states can
be said to be cold or hot relative to each other. Another (less
preferred by us) approach to cooling would be to employ global measures such as
entropy.  }. 

For simplicity we assume that {\bf S} is a two-level system with
energies \footnote{Consider a multi-level system whose energy gap
between the ground-state and the first excited state is smaller than the
gap between the first and second excited state. For sufficiently low
initial temperatures this system can be regarded as effectively
two-level system, because populations of the second and higher energy
levels can be neglected.}
\BEA
0\,\leq\,\epsilon,
\EEA
and equilibrium probabilities
\BEA
\label{stuks}
p_0 \equiv \langle 0|\rho_{\rm \bf S}(0)|0\rangle
= \frac{1}{1+ e^{-\beta \epsilon}}, \qquad p_1=\frac{e^{-\beta \epsilon}}{1+e^{-\beta \epsilon}}.
\EEA
The reservoir $\R$ is a $M$-level system with energies
\BEA
0=\mu_0\,\leq\,\mu_1 \,\leq\,\ldots \,\leq\, \mu_{M-1}\equiv \mu,
\label{goelro}
\label{zor}
\EEA
and initial, equilibrium probabilities [see (\ref{mu}, \ref{brams})]
\BEA
\label{lenin}
\pi_l=\frac{e^{-\beta \mu_l}}{1+\sum_{k=1}^{M-1} e^{-\beta \mu_k}  }, \qquad l=0,\ldots,M-1.
\EEA

The eigenvalues of the initial density matrix $\rho_{\S+\R}$ read
\begin{gather}
\label{in}
\{\omega_k\}_{k=0}^{2M-1}
=(p_0\pi_0,\, p_1\pi_0,\, p_0\pi_1,\, p_1\pi_1,\ldots).
\end{gather}

\subsection{Unattainability of the absolute zero}
\label{nozero}

The above set-up suffices for showing the unattainability of the
absolute zero, $p_0(\tau)>1$, for the target of cooling {\bf S}, given
the initial state (\ref{mu}) of the reservoir. The final ground-state
probability $p_0(\tau)$ of {\bf S} follows from (\ref{mu}--\ref{lenin})
\BEA
\label{stalin}
1-p_0(\tau)= {\sum}_{i=0}^1{\sum}_{\alpha,\gamma=0}^{M-1} p_i\pi_\gamma
[1-|\langle 0,\alpha|U|i,\gamma   \rangle|^2],
\EEA
where $\{|i\rangle\}_{i=0}^1$ and $\{|\alpha\rangle\}_{i=0}^{M-1}$ are
the eigenbases of, respectively, $H_{\rm \bf S}$ and $H_{\rm \bf R}$ in
(\ref{brams}), and $\{\pi_\alpha\}_{\alpha=0}^{M-1}$ is given by
(\ref{lenin}). Since ${\sum}_{i=0}^1{\sum}_{\gamma=0}^{M-1}|\langle 0,\alpha|U|i,\gamma   \rangle|^2=1$, some inequlities
$1\geq |\langle 0,\alpha|U|i,\gamma \rangle|^2$ must be strict. Noting also
$p_i\pi_\gamma>0$ we deduce from (\ref{stalin})
\BEA
\label{molotov}
p_0(\tau)< 1 \qquad {\rm (strict~~ inequality)}.
\EEA
This argument applies for $M=\infty$, where, e.g.,
${\sum}_{i=0}^1{\sum}_{\gamma=0}^{\infty}|\langle 0,\alpha|U|i,\gamma
\rangle|^2=1$ and ${\sum}_{\gamma=0}^{\infty}\pi_\gamma=1$ are
convergent series, and where $U$ is unitary for $M=\infty$. Note that
the argument leading to (\ref{molotov}) is essentially based on
$p_i\pi_\gamma>0$ and hence $\pi_\gamma>0$; see (\ref{lenin}). 

With trivial changes the derivation of (\ref{molotov})
applies for a multi-level system {\bf S}.

\subsection{Lower bounds for work}

Here we summarize restrictions imposed by the second law on the work
needed for cooling. Our presentation uses the same ideas and methods as 
\cite{lindblad,passivity}; see also \cite{parrondo} in this context. 

Recall that for the considered unitary (thermally isolated) process the
work done on the system is equal to the [average] energy difference
\cite{lindblad,passivity}:
\BEA
\label{worko}
W={\rm tr}(\,[\,\rho_{\rm \bf S+R}(\tau)
-\rho_{\rm \bf S+R}\,]H_{\rm \bf S+R}),
\EEA
where we recall that the interaction $V(t)$ with external fields is
switched off after the final time; see (\ref{ga}).  

Since the initial state (\ref{mu}) is at a Gibbsian equilibrium, any
unitary operator that changes this state costs some work. Indeed, given
the initial state (\ref{mu}) and the unitary dynamics implemented by a
cyclically changing Hamiltonian (\ref{ga}, \ref{fio}), the work
(\ref{worko}) invested in {\bf S+R} amounts to
\BEA
\label{kosa1}
\beta W&&= {\rm tr} (\rho_{\rm \bf S+R}\ln \rho_{\rm \bf S+R}-\rho_{\rm \bf S+R}(\tau)\ln \rho_{\rm \bf S+R})\\
\label{kosa2}
&&= {\rm tr}
(\rho_{\rm \bf S+R}(\tau)\ln \rho_{\rm \bf S+R}(\tau)-\rho_{\rm \bf S+R}(\tau)\ln \rho_{\rm \bf S+R}) ~~~~\\
&& \equiv S[\rho_{\rm \bf S+R}(\tau)||\rho_{\rm \bf S+R}]\geq 0,  
\label{kosa3}
\EEA
where in moving from (\ref{kosa1}) to (\ref{kosa2}) we used the
unitarity of dynamics; see (\ref{fio}). The relative entropy
$S[\rho_{\rm \bf S+R}(\tau)||\rho_{\rm \bf S+R}]$ is non-negative and
nullifies only if $\rho_{\rm \bf S+R}(\tau)=\rho_{\rm \bf S+R}$; see
Refs.~\cite{ve} for further features of the relative entropy and its
role in (quantum) information theory. Thus any change of the initially
equilibrium state via a cyclic Hamiltonian process costs some work. This
work is also a resource of dynamic cooling. 

In the context of cooling one can derive more stringent lower bounds on
the work. Using (\ref{brams}) we rearrange (\ref{kosa2}):
\begin{gather}
\label{mega}
\beta W= S[\rho_{\rm \bf S}(\tau)||\rho_{\rm \bf S}]+S[\rho_{\rm \bf R}(\tau)||\rho_{\rm \bf R}]+I_{\rm \bf SR}(\tau),\\
I_{\rm \bf S+R}(\tau)\equiv
{\rm tr} [\,\rho_{\rm \bf S+R}(\tau)\ln \rho_{\rm \bf S+R}(\tau)\,]\nonumber\\
-{\rm tr}[\,\rho_{\rm \bf S}(\tau)\ln \rho_{\rm \bf S}(\tau)\,]-
{\rm tr} [\,\rho_{\rm \bf R}(\tau)\ln \rho_{\rm \bf R}(\tau)\,]\geq 0.
\label{osho}
\end{gather}
$I_{\rm \bf S+R}(\tau)$ is the mutual information between ${\rm \bf S}$ and ${\rm \bf R}$ in the final state.
$I_{\rm \bf S+R}(\tau)\geq 0$ due to the sub-additivity of the entropy. Now (\ref{mu}, \ref{brams}, \ref{mega}) and 
(\ref{kosa3}, \ref{osho}) produce
\BEA
W &\geq& TS[\rho_{\rm \bf S}(\tau)||\rho_{\rm \bf S}] \nonumber \\
  &=& F[\rho_{\rm \bf S}(\tau)] - F[\rho_{\rm \bf S}]\equiv \Delta F\geq 0,
\label{gaspar}
\EEA
where for any density matrix $\sigma$, the free energy at the initial temperature $T$ is defined as 
\BEA
\label{gasparian}
F[\sigma]\equiv {\rm tr}(H_{\rm \bf S}\sigma)+T{\rm tr}(\sigma\ln\sigma).
\EEA
If $\sigma$ is a Gibbsian density matrix at temperature $T$, $F[\sigma]$
coincides with equilibrium free energy [cf. (\ref{buka})].  The meaning
of (\ref{mega}) is that the work equals the sum of the free energies
differences for {\bf S} and {\bf R} + the energy stored in the mutual
information between {\bf S} and {\bf R}. 

$\Delta F$ in (\ref{gaspar}) is different from the free energy
difference (\ref{a3}) of the Nernst set-up, where both the initial and
final states are at equilibrium. Hence in (\ref{a3}) the temperatures
$T_{\rm in}$ and $T_{\rm fin}$ refer to the initial and final states,
respectively. For dynamic cooling only the initial states are at
Gibbsian equilibrium (\ref{mu}). Hence the free energy in (\ref{gaspar},
\ref{gasparian}) refers to the initial temperature only. To compare
quantitatively (\ref{gaspar}) with (\ref{a3}), we note that $g_{\rm in}$
in (\ref{a3}) refers within dynamic cooling to the {\bf S--R}
(system-reservoir) interaction that is zero both initially and finally.
Assume that the final state $\rho_{\rm \bf S}(\tau)$ of {\bf S} is
Gibbsian (canonic equilibrium) at temperature $T_{\rm fin}$. Then the
thermodynamic entropy in (\ref{buka}) is the von Neumann entropy from
(\ref{gasparian}). Recalling that $T_{\rm fin}<T_{\rm in}\equiv T$ we
obtain from (\ref{a3}, \ref{gaspar})
\BEA
F[T_{\rm fin}, g_{\rm in}]&-& F[T_{\rm in}, g_{\rm in}]-
TS[\rho_{\rm \bf S}(\tau)||\rho_{\rm \bf S}] \nonumber\\
&=&-{\rm tr}[\rho_{\rm \bf S}(\tau)\ln\rho_{\rm \bf S}(\tau)]\,
(T_{\rm fin}-T_{\rm in})\geq 0.~~~
\label{orda}
\EEA
Thus the difference between (\ref{gaspar}) and (\ref{a3}) tends to disappear with
the final entropy of {\bf S}. Such a situation will be met in section \ref{N}.

\subsection{Summary of conditions }
\label{comparo}

In  analogy to section \ref{summa_contra} we summarize the dynamical cooling set-up in terms of the following task list:

{\bf D1.} Initial state: canonical equilibrium system and reservoir; they are uncoupled.

{\bf D2.} System (target of cooling): microscopic.

{\bf D3.} Cooling dynamics: unitary in the space of the system + reservoir. Not necessarily quasi-equilibrium.

{\bf D4.} Reservoir: possibly microscopic. Not necessarily larger than the system.

{\bf D5.} System-reservoir interaction (driven by external field): changes cyclically. 

{\bf D2} makes obsolete the need for the entropic formulation of the
third law, since a thermally isolated process done on a finite {\bf S}
is not uniquely characterized by its entropy \cite{abn} (instead the
full spectrum of the density matrix determines the set of states that
can be reached by unitary processes). In this paper we assumed ${\bf
D2}^*$ rather than {\bf D2}, i.e. that the system {\bf S} is
two-dimensional. This more restrictive condition should be relaxed in
future studies. {\bf D5} is motivated in the same way as {\bf N5}. 
To grasp the difference between {\bf N4} and {\bf D4},
we shall study in section \ref{N} the dynamic cooling set-up under condition {\bf N4}.

As for the tractability we keep the conditions {\bf T1--T3} from section
\ref{summa_contra}.  The unattainability of the absolute zero is proven
in section \ref{nozero} via {\bf D1} and {\bf D3} |without explicit
reference to {\bf T1--T3}. Nevertheless, {\bf T2} will be needed below
for understanding limitations on reaching lowest (non-zero)
temperatures, while {\bf T3} is inherently demanded for applications of
the dynamical cooling: The cycle time now refers to the time needed to
realize the unitary operator (\ref{ga}) via a suitable Hamiltonian. 

\section{Optimal Reservoir}
\label{opto}

The task list of section \ref{comparo} does not yet fully specify the
reservoir.  In this section and the following two we will investigate
three different models; their respective parameters are then shown to
characterize the way in which cooling can be achieved. 

\subsection{Max-Min cooling scenario}
\label{maxmin}

The purpose of dynamic cooling is to increase the ground state
probability of {\bf S} subject to the constraints listed in section
\ref{comparo}. Obviously, some set-ups will work better for the desired
task than others. For studying principle limitations we should thus
optimize the design, i.e. try to maximize this very probability. This
has to be done {\it i)} over all unitary transformations $U$ in
(\ref{fio}); {\it ii)} over the energy level distribution of the
reservoir assuming that they are bound from above, i.e., $\mu$ in
(\ref{zor}) is a finite, fixed number. This parameter of the upper
energy level of the reservoir will turn out below to control dynamic
cooling. We denote these maximization strategies as ${\rm max}_{U}$ and
${\rm max}_{\mu_k}$ respectively. The overall maximization is ${\rm
max}_{U,\mu_k}$. 

Generally, the above maximization procedure will specify a set of
equivalent scenarios only, i.e. leave some free parameters (see below).
Additional requirements may be imposed to reduce this ambiguity.  Here
we attempt to minimize the work needed for given cooling effect.  This
is why our intended goal requires a "max-min" scenario: first the ground
state probability of $\S$ is maximized and only after that the work is
minimized. 

\subsection{Maximal cooling}
\label{maxo}

Assume that $\R$ has an even number of energy levels:
\BEA
M=2n.
\EEA
Appendix \ref{apo} shows that the unitary operator, which leads to the
largest final ground-state probability for $\S$, amounts to permuting
the elements (\ref{in}) of the initial density matrix $\rho_{\bf S+R}$.
We recall that it is diagonal in the energy representation. Hence in the
maximal cooling regime the final state of the two-level target is
diagonal in the energy representation: it has a well-defined
temperature, which is hence not imposed, but emerges out of
optimization. 

The fact that the two-level target of cooling ends up in an energy
diagonal density matrix (with the ground-level probability greater than
the excited level probability) is straightforward to establish:
otherwise, there will be a unitary operator acting only on the
two-level system such that its ground-state probability is increased. 

Consider the eigenvalue vector of the final state of {\bf S+R} with the
largest ground-state probability $p_0(\tau)$ of {\bf S}.  In this vector
the largest $2n$ elements of the vector (\ref{in}) are at the odd places
[counting starts with $1$], and the final, maximized ground-state
probability for $\S$ reads
\BEA
\label{cobra}
{\rm max}_U[\,p_0(\tau)\,] = {\bf smax}_{2n}[\, \{\omega_k\}_{k=0}^{4n-1}\,],
\EEA
where ${\bf smax}_{k}[{\bf a}]$ returns the sum of $k$ largest elements
of vector ${\bf a}$. Finding the $2n$ maximal elements of $\{\omega_k \}_{k=0}^{4n-1}$
leaves some freedom in $U$. One represents (\ref{cobra}) as
\BEA
{\rm max}_U[\,p_0(\tau)\,] = p_0{\sum}_{k=0}^{n-1}\pi_k +p_1\pi_0~~~~~~~~~~~~~~~~~~\nonumber\\
+{\bf smax}_{n-1}[\,p_1\pi_1,\ldots,p_1\pi_{n-1},p_0\pi_n,\ldots,p_0\pi_{2n-2}     \,],
\label{rashid}
\EEA
where we used (\ref{in}) and straightforward induction over $n$.

Maximizing (\ref{cobra}) over the energy levels (\ref{goelro}) 
amounts to maximizing it over the Boltzmann weights 
$v_k=e^{-\beta\mu_k}$ [see (\ref{goelro}, \ref{lenin})] under
constraints
\BEA
1\geq v_1\geq...\geq v_{2n-1}\geq v=e^{-\beta\mu}.
\EEA
Since the LHS of (\ref{cobra}) is a ratio of linear functions of $v_k$,
it can maximize only at the borders of the allowed range of $v_k$, i.e.,
some of $v_k$'s are equal to $1$, while others are equal to $v$. The physical meaning of
this result is that {\it the optimal reservoir is an effective two-level
system}, a fact that greatly simplifies searching for the maximal
(final) ground-state probability of {\bf S}.  To illustrate this result
consider an example in (\ref{rashid}): $n=2$, ${\bf
smax}_{1}[p_1\pi_1,p_0\pi_2]={\rm max}[p_1\pi_1,p_0\pi_2]=p_1\pi_1$.
This reduces (\ref{rashid}) to $\pi_0+\pi_1$, which maximizes for
$v_1=1,v_2=v$. 

It now remains to check all possible arrangements of energy levels that
render {\bf R} an effectively two-level system. This check produces the
maximum of (\ref{rashid}) for
\BEA
\label{bobo}
\mu_0=\ldots=\mu_{n-1}=0, \qquad \mu_{n}=\ldots=\mu_{2n-1}=\mu.
\EEA
Hence the optimal reservoir for cooling a two-level system has to have a degenerate ground-state for $n\geq 2$.
The maximal ground-state probability now reads
\BEA
\label{dardibun}
{\rm max}_{U,\mu_k}[\,p_0(\tau)\,]&=& {\sum}_{k=0}^{n-1} \pi_k=\frac{1}{1+e^{-\beta \mu}}\\
&=&\frac{1}{1+e^{-\epsilon/ T_{\rm min}}},  \nonumber
\EEA
where 
\BEA
\label{rer}
T_{\rm min} = {T\epsilon}/{\mu}. 
\EEA
Thus (\ref{dardibun}) is the maximal probability for a reservoir with maximal energy $\mu$; 
$T_{\rm min}$ is the minimal temperature. Note that (\ref{rer}) is consistent with 
the unattainability argument (\ref{molotov}). For $\mu \rightarrow  \infty$, $T_{\rm min}$ would approach zero.

We get cooling, i.e.  ${\rm max}_{U,\mu_k}[\,p_0(\tau)\,] >p_0$,
only when the largest energy of {\bf R} is larger than the 
energy of {\bf S}:
\BEA
\label{vajvaj}
\mu>\epsilon.
\EEA
This asymmetry ensures that the two-level is cooled in presence of the reservoir.

\subsection{Minimal work}
\label{mino}

We turn to calculating the minimal work given the maximal probability (\ref{dardibun}). 
Write down (\ref{in}) under (\ref{bobo}):
\BEA
\underbrace{\underline{p_0\pi}, p_1\pi,\ldots, \underline{p_0\pi}, p_1\pi}_{2n\,{\rm elements}}, 
\underbrace{p_0\tilde{\pi}, \underline{p_1\tilde{\pi}},\ldots, p_0\tilde{\pi},\underline{p_1\tilde{\pi}}}_{2n\,{\rm elements}},
\label{gov}
\EEA
where we defined $\pi=\frac{1}{n(1+v)}$ and $\tilde{\pi}=\frac{v}{n(1+v)}$.

Eq.~(\ref{dardibun}) is found after the first $2n$ elements in
(\ref{gov}) are distributed over the odd places in the eigenvalue list
of $\rho_{\rm \bf S+R}^{\rm fin}$ [counting starts from 1]. Concrete
places they occupy are not important for obtaining the optimal cooling
effect (\ref{dardibun}): Those various orderings constitute an
equivalence class.  Concrete places, however, become important for
minimizing the work. Note that for a fixed final ground-state
probability of the two-level system minimizing the work amounts to
maximizing the overall probability of the lowest level
$\mu_0=\ldots=\mu_{n-1}=0$ for the final state of the reservoir; see
(\ref{bobo}). Then (\ref{gov}) implies that the minimal work is obtained
for a permutation that does not touch the underlined elements in
(\ref{gov}), but interchanges those that are not underlined: each
$p_1\pi$ with some element $p_1\tilde{\pi}$; see Appendix \ref{apo} for
the argument reducing the considered unitary to a permutation. Recalling
that the initial energy of the system+reservoir is $\frac{\mu
e^{-\beta\mu}}{1+e^{-\beta\mu}}+ p_1\epsilon$ and using (\ref{worko}) we
get
\BEA
W=(\mu-\epsilon)\left[\frac{1}{1+e^{-\beta\mu}}-p_0\right],
\label{trdao}
\EEA
for the work. Now (\ref{trdao}) shows that reaching the lowest possible
temperature $T_{\rm min}$ requires the work 
\BEA
\label{twostar}
W \simeq \mu p_1 \sim 1/T_{\rm min}.
\EEA
This constitutes a parametrization of the attainability constraints for
dynamical cooling: Work $W$ and and the inverse temperature reached,
$1/T_{\rm min}$, are proportional. Accepting that $W$ cannot be infinite (see
condition {\bf T2} in section \ref{summa_contra} and \ref{comparo}), we
have to accept that the minimal temperature reached cannot be zero. 

Let now the number of energy levels of {\bf R} be an odd number: $M=2n+1$.
Instead of (\ref{rashid}) we get
\BEA
{\rm max}_{U}[\,p_0(\tau)\,]
= p_0{\sum}_{k=0}^{n-1}\pi_k +p_1\pi_0\nonumber~~~~~~~~~~~~~~~~~\\
+{\bf smax}_{n}[\,p_1\pi_1,\ldots,p_1\pi_{n-1},p_0\pi_n,\ldots,p_0\pi_{2n-1}     \,].
\label{rashido}
\EEA
The final ground state probability is maximized for 
\BEA
\label{boboo}
\mu_0=\ldots=\mu_{n-1}=0, \qquad \mu_{n}=\ldots=\mu_{2n}=\mu,
\EEA
meaning again that the optimal reservoir has to have a degenerate ground-state for $n\geq 2$.
The maximal ground state probability reads:
\BEA
\label{dok}
{\rm max}_{U,\mu_k}[\,p_0(\tau)\,]=\frac{n+p_0 e^{-\beta\mu}}{n+(n+1)e^{-\beta\mu}}.
\EEA
Condition (\ref{vajvaj}) is still needed, and the
qualitative conclusion from studying the work cost is the same as above. 

In constrast to (\ref{dardibun}), expression (\ref{dok}) already depends
on the initial probability $p_0$ of {\bf S}. Eqs.~(\ref{dardibun}) is
recovered from (\ref{dok}) for a many-level reservoir $n\gg 1$. Instead
of assuming such a many level reservoir with the optimal unitary
operating on the joint Hilbert space of this reservoir and the target
system, we can relate (\ref{dok}) to (\ref{dardibun}) under a weaker condition.  Apply the
cooling protocol repeatedly with the reservoir|having a finite, odd
number of energy levels|re-prepared in its equilibrium state, e.g., via
fast relaxation, as it happens with electronic spins in solid state NMR
\cite{abo}. Then the ground-state probability increases iteratively as
[see (\ref{dok})]:
\BEA
p_0^{[l+1]}=\frac{n+p_0^{[l]}
e^{-\beta\mu}}{n+(n+1)e^{-\beta\mu}}, \qquad l=1,2,\ldots. 
\EEA
For $l\gg 1$
the result of this iteration converges to (\ref{dardibun}).

\subsection{Trade-off between maximizing cooling and minimizing work}
\label{trade-off}

The existence of the above work cost for dynamic cooling raises the
following question: to what extent can we reduce this cost if,
given the upper bound on the reservoir energy spectrum [see
(\ref{zor})], we sacrifice some ground-state probability of the target
system, i.e., instead of reaching (\ref{dardibun}) we agree to reach a
somewhat lower final probability. As compared to (\ref{dardibun}), this 
will require a different reservoir and a different unitary transformation
for cooling. The answer to this question is that even a small decrease in the achieved 
ground-state probability can significantly reduce the work-cost. We shall 
illustrate this fact via an example. 

Let us take $M=2n=4$, and postulate a unitary operator that permutes 
the eigenvalues of the initial state (\ref{in}) so that the final-state eigenvalues read
\BEA
\label{fi}
(p_0\pi_0,\, p_0\pi_2,\, p_1\pi_0,\, p_1\pi_2,\, p_0\pi_1,\, p_0\pi_3,\, p_1\pi_1,\, p_1\pi_3).
\EEA
The logics of obtaining (\ref{fi}) from (\ref{in}) is as follows. The
first four probabilities $p_0\pi_0,\, p_1\pi_0,\, p_0\pi_1,\, p_1\pi_1$ in (\ref{in})
are moved to odd places producing
\BEA
\label{1}
\hat{p}_0(\tau)=\pi_0+\pi_1= \frac{1+e^{-\beta \mu_1}}{1+ e^{-\beta \mu_1} +e^{-\beta \mu_2}+e^{-\beta \mu_3}  } ,
\EEA
which is similar to $\pi_0+\pi_1$ in (\ref{dardibun}). The remaining
four probabilities $p_0\pi_2,\, p_1\pi_2,\, p_0\pi_3,\, p_1\pi_3$ [last
four elements in (\ref{in})] are then arranged in between without
changing their mutual order.  The work and the final probabilities for
the four reservoir energy levels read
\BEA
\label{2}
\label{4.0}
&&W=\epsilon (p_0-\hat{p}_0(\tau)) + {\sum}_{k=1}^3\mu_k (\pi_k(\tau)-\pi_k),\\
\label{4.00}
&&\pi_0(\tau)=p_0(\pi_0+\pi_2), \qquad \pi_1(\tau)=p_1(\pi_0+\pi_2), \\
&&\pi_2(\tau)=p_0(\pi_1+\pi_3), \qquad \pi_3(\tau)=p_1(\pi_1+\pi_3).
\label{4.000}
\label{3}
\EEA
Let us assume that
\BEA
\label{ole}
\beta\mu_2=\beta\mu_3\equiv \beta\mu \gg 1
\EEA 
is a fixed number. Hence $\mu_1$ is the only parameter 
over which we can minimize (\ref{4.0}). In the limit $\beta\mu \to \infty$ 
both (\ref{dardibun}) and (\ref{1}) converge to $1$:
\BEA
\label{akino}
{\rm max}_{U,\mu_k}[\,p_0(\tau)\,]\to 1, \qquad \hat{p}_0(\tau)\to 1, 
\EEA
while their ratio is a finite number
\BEA
\label{odzin}
\frac{1-{\rm max}_{U,\mu_k}[\,p_0(\tau)\,]}{1-\hat{p}_0(\tau)}=\frac{1}{2}(1+e^{-\beta\mu_1})+{\cal O}[e^{-\beta\mu}],
\EEA
where ${\cal O}[e^{-\beta\mu}]$ can be neglected due to (\ref{ole}). 

We now show that although (\ref{dardibun}) and (\ref{1}) are quite close to each other,
the work needed to obtain (\ref{1}) deviates significantly from (\ref{trdao}). Note that
$\pi_2$ and $\pi_3$ are exponentially small with $\beta\mu \to \infty$. We neglect such 
terms in (\ref{4.0}) and obtain
\BEA
\label{wo1}
W=-\epsilon p_1 +\mu_1(p_1\pi_0-\pi_1)+\mu \pi_1, \\
\beta W=p_1\ln\frac{p_1}{1-p_1}+\frac{\beta\mu_1(p_1-e^{-\beta\mu_1})+\beta\mu e^{-\beta\mu_1}}{1+e^{-\beta\mu_1}}.
\label{blo}
\EEA
To get (\ref{blo}, \ref{wo1}) from (\ref{4.0}--\ref{4.000}) we employed (\ref{stuks}, \ref{lenin}).  
It should now be clear that for $\beta\mu\gg 1$ we get a logarithmically growing work, $\beta W=p_1\ln
[\frac{p_1\beta\mu}{1-p_1}]$, if we choose $\beta\mu_1=\ln(\beta\mu)$.
The minimization of the RHS of (\ref{blo}) over $\beta\mu_1$ satisfying
$\mu_1\leq \mu$ produces a very similar result. For $\beta\mu\gg 1$ 
neglect in (\ref{blo}) all terms with $e^{-\beta\mu_1}$ except the factor $\beta\mu e^{-\beta\mu_1}$,
where a small term $e^{-\beta\mu_1}$ is multiplied by the large $\beta\mu$. Now substitute 
$\beta\mu_1=\ln\left[\frac{\beta\mu}{yp_1}\right]$ into (\ref{blo}), where $y$ is a new variable. This produces
$\beta W =p_1 \ln\left[\frac{p_1}{1-p_1}\right]+p_1(y+\ln\left[\frac{\beta\mu}{yp_1}\right])$.
After differentiating over $y$ we obtain for the minimum: $y=1$, or 
\BEA
\beta W =p_1 \ln\left[\frac{\beta\mu\, e}{1-p_1}\right].
\label{mlo}
\EEA
and thus a logarithmic rather than linear increase of $W$ with $\mu$, i.e. with $\beta'$. In the limit
$\beta\mu \rightarrow \infty$ the corresponding ground state probability is only slightly reduced:
\BEA
\label{oster}
\frac{1-{\rm max}_{U,\mu_k}[\,p_0(\tau)\,]}{1-\hat{p}_0(\tau)}=\frac{1}{2}\left(1+\frac{p_1}{\beta\mu}\right).
\EEA
This equation can be rewritten in terms of temperatures. Recall (\ref{akino}, \ref{dardibun}), 
and introduce a temperature $\hat{T}$ via $\hat{p}_0(\tau)=(1+e^{-\epsilon/ \hat{T}})^{-1}$. 
Then (\ref{oster}) reads
\BEA
\frac{1}{\hat{T}} -\frac{1}{T_{\rm min}}= \frac{1}{\epsilon}\ln\left[ \frac{1}{2}\left(1+\frac{p_1}{\beta\mu}\right) \right].
\EEA
For $\beta\mu\to\infty$ the r.h.s. of this equation goes to $-\frac{\ln 2}{\epsilon}$.

If one allows further deviations from the maximal probability
(\ref{dardibun}) [more than (\ref{oster})], then it is possible to
reduce the work even below the logarithmic dependence (\ref{mlo}). Appendix
\ref{apa} shows that the example (\ref{fi}) generalized to $n=4$
(eight-level reservoir) produces [for $\beta\mu\gg 1$] the
double-logarithmic scaling of the minimimal work, $W\simeq \ln [\ln
[\beta\mu]]$ provided that the deviation from the maximal probability
scales as 
\BEA
\label{goris}
\frac{1-{\rm
max}_{U,\mu_k}[\,p_0(\tau)\,]}{1-\hat{p}_0(\tau)}=\frac{1}{4}(1+{\cal O}
\left[\frac{1}{\ln(\beta\mu)}\right]). 
\EEA

Realistic reservoirs tend to deviate from the optimal model studied above. In the following we
investigate models subject to additional constraints.

\comment{Note the following difference between the fully optimal reservoir
(\ref{boboo}) and the work-compromised reservoir (\ref{ole},
\ref{dorosh}) studied above. The former cools the two-level system to
the same final ground-state probability (\ref{dok}) independently from
the initial state of the two-level system [provided that (\ref{vajvaj})
holds]. This is similar to most ordinary refrigerators.  The structure
of the work-compromised reservoir already depends on the initial state
of the two-level system, as (\ref{dorosh}) shows. }

\section{Homogeneous spectrum of reservoir}
\label{homogeneous}

\subsection{Finite number of levels}

We saw above that the maximal energy $\mu$ [see (\ref{zor})] is an
essential parameter for dynamical cooling via a reservoir with a finite
number of energy levels: cooling to ground-state probability $1$ is
possible only for $\mu\to\infty$. In the optimal scenario
(\ref{dardibun}) (with two levels only) $\mu$ conincides with the energy
gap. 

Here we study the simplest cooling scenario, where $\mu$ can be large
without increasing the level spacing. This scenario also illustrates
the limit of the inifnite-dimensional Hlibert space for the reservoir. 

We assume that {\bf R} has $M$ equidistant
energy levels:
\BEA
\label{maya}
\mu_k=\delta k,\qquad k=0,1,\ldots,M-1,
\EEA
where $\delta>0$ is the reservoir energy gap. 
We realize with respect to this reservoir the same max-min scenario. 
However, no optimization over the energy gaps of the
reservoir will be carried out, i.e., $\delta$ is a fixed parameter. 

The initial-state eigenvalues of {\bf S+R} read:
\BEA
\label{baum}
\frac{1}{Z}[p_0,p_1,p_0v,p_1v, p_0v^2,p_1v^2,\ldots,p_0v^{M-1},p_1v^{M-1}],
\EEA
where we defined
\BEA
v=e^{-\beta\delta}, \qquad Z={\sum}_{k=0}^{M-1}v^k=(1-v^M)/(1-v).
\EEA
We define $\alpha$ from
\BEA
p_0v^\alpha\geq p_1\geq p_0v^{\alpha+1}, \quad \alpha=\left\lfloor \frac{\ln (p_0/p_1)}{\ln(1/v)}   \right\rfloor
=\left\lfloor\frac{\epsilon}{\delta}   \right\rfloor,
\label{mudr}
\EEA
where $\lfloor x \rfloor$ is the floor integer part of $x$, e.g., $\lfloor 0.99 \rfloor=0$.
Let us re-order the elements of vector (\ref{baum}) (first the largest element, then next to the largest, etc):
\begin{gather}
\label{komar1}
\vec{\omega}=
\frac{1}{Z}[\,
\underbrace{p_0,p_0v,\ldots,p_0v^\alpha}_{\alpha+1}, \\
\label{komar2}
\underbrace{p_1,p_0v^{\alpha+1},p_1v,p_0v^{\alpha+2}, \ldots,
p_1v^{M-\alpha-2},p_0v^{M-1}}_{2(M-\alpha-1)}, \\
\underbrace{p_1v^{M-\alpha-1}, p_1v^{M-\alpha},\ldots, p_1v^{M-1}}_{\alpha+1}
\,],
\label{komar3}
\end{gather}
where the curly bracket shows the number of elements in each
group. In (\ref{komar1}--\ref{komar3}) we select the first $M$ elements; their
sum will give the maximal final ground-state
probability of the target two-level system. Then the optimal unitary transformation amounts
to distributing those $M$ largest elements over the odd places in the
eigenvalue list of the final density matrix. Clearly, $M=\alpha+1$ means 
that no cooling is possible with the considered reservoir.

Let $M-(\alpha+1)$ be an even number:
\BEA
\label{basta}
M-(\alpha+1) =2 s,
\EEA
where $s$ is an integer. Now (\ref{komar2}) is factorized as
\BEA
\label{apara}
(\ref{komar2})=\underbrace{p_1,p_0v^{\alpha+1},\ldots,p_1v^{s-1},p_0v^{\alpha+s}}_{M-\alpha-1},\nonumber\\
\underbrace{p_1v^s,p_0v^{\alpha+s+1},\ldots,p_1v^{M-\alpha-2},p_0v^{M-1}}_{M-\alpha-1}
\EEA
Thus the final ground-state probability of the target two-level system reads
\footnote{Likewise, for an odd $M-(\alpha+1)$ we introduce
$M-\alpha-1 =2 u+1$, where $u$ is an integer. Instead of (\ref{sad}) we get
${\rm max}_U[\,p_0(\tau)\,]=\frac{1-p_1v^{u+1}-p_0v^{\alpha+u+1}}{1-v^M}$,
which implies the same consequence (\ref{saba}).}
\BEA
{\rm max}_U[\,p_0(\tau)\,]
=\frac{1}{Z}[\,p_0{\sum}_{k=0}^{\alpha+s}v^k+p_1{\sum}_{k=0}^{s-1}v^k\,]\nonumber\\
=\frac{1-p_1v^s-p_0v^{\alpha+s+1}}{1-v^M}.
\label{sad}
\EEA
As expected, (\ref{sad}) is smaller than the bound (\ref{dardibun}) with $\mu=\delta(M-1)$ .

\subsection{Infinite number of levels}
\label{horror}

For finite $\delta$ and $\alpha$ we get from (\ref{sad}):
\BEA
{\rm max}_U[\,p_0(\tau)\,]\to 1 \quad {\rm when} \quad M\to\infty ~~{\rm and}~~ s\to\infty.
\label{saba}
\EEA

The minimal work necessary for (\ref{saba})
is {\it finite}. In showing this let us restrict ourselves with the simplest case 
$\alpha=0$ in (\ref{basta}). Eqs.~(\ref{komar1}--\ref{komar3}) now read
\BEA
\underbrace{p_0,p_1,\ldots,p_1v^{(M-3)/2},p_0v^{(M-1)/2}}_{M},\nonumber\\
\underbrace{p_1v^{(M-1)/2},p_0v^{(M+1)/2},\ldots,p_0v^{M-1},p_1v^{M-1}}_{M}
\label{gus}
\EEA
Eq.~(\ref{gus}) implies for the minimal work in the limit (\ref{saba})
\begin{gather}
W=\epsilon (p_0-p_0(\tau))+\frac{\delta}{Z}(p_1+p_0v^{\frac{M+1}{2}}) \sum_{k=0}^{(M-3)/2}(2k+1)v^{k}\nonumber\\
~~~+\frac{\delta}{Z}(p_0v+p_1v^{\frac{M+1}{2}}) \sum_{k=0}^{(M-3)/2}(2k+2)v^{k}-\frac{\delta}{Z}\sum_{k=0}^{M-1}k v^k\nonumber\\
=(\delta-\epsilon)p_1+\frac{\delta v}{1-v},~~~~~~~~~~~~~~~~~~~~~~~~~~~~~~~
\label{glasgow}
\end{gather}
which is clearly finite, positive and is larger than the bound $\Delta F=-\frac{1}{\beta}\ln(1+e^{-\beta\epsilon})$ demanded by
(\ref{gaspar}).

The convergence (\ref{saba}) does not mean that for
$M=\infty$ (harmonic oscillator spectrum) we would get ${\rm
max}_U[\,p_0(\tau)\,]=1$ (or $T_{\rm min}=0$), because for $M=\infty$
the transformation that leads to (\ref{sad}) is not even bijective, let
alone unitary (i.e., the limit of unitary processes for $M\to\infty$ is
not unitary); see in this context our discussion after (\ref{molotov}).
Let us take $\alpha=0$ in (\ref{komar1}--\ref{komar3}). For a
finite $M$ the optimal unitary|which we recall amounts to a
permutation|distributes the first half of the vector
(\ref{komar1}--\ref{komar3}) over the odd places in the final vector.
The elements from the second half are distributed into even places of
the final vector. For $M=\infty$ this second (sub)process disappears in
infinity. Hence the limit of the above permutation for $M\to\infty$ is
not even bijective. 

Cooling with the harmonic oscillator reservoir ($M=\infty$) is a
well-defined problem provided that one ensures that the operator $U$
stays unitary for $M=\infty$. For each such unitary we have
$p_0(\tau)<1$; see (\ref{molotov}). But the maximum ${\rm
max}_U[\,p_0(\tau)\,]$ now does not exist. It can be substituted by
supremum 
\BEA
\label{supo}
{\rm sup}_U[\,p_0(\tau)\,]=1, 
\EEA
as follows from (\ref{saba}). Eq.~(\ref{supo}) points out on an
important difference between the infinite and 
finite-level situation. 

\section{Reservoir consisting of $N$ identical spins.}
\label{N}

We study this modular case for three reasons: First, in the {\it
thermodynamic limit} $N\to\infty$ the reservoir will become a standard
thermal bath; hence one expects to establish connections between the
Nernst set-up [section \ref{opera}] and the dynamic cooling set-up
[section \ref{setup}]. Second, the model is relevant for polarization
transfer experiments, and thus was widely studied|albeit for the
high-temperature limit only|in the NMR literature \cite{ole,schulman}. 

\subsection{Maximal cooling}

Now reservoir {\bf R} consists of $N$ identical spins, each one with
energies $0$ and $\delta>0$. The energy levels 
of the reservoir, $E_A = \delta A$, are equidistant with degeneracy
\BEA
\label{sos}
d_A=\frac{N!}{(N-A)!\,A!}.
\EEA
One can apply (\ref{komar1}--\ref{komar3}) to this situation|with
$\alpha$ being defined as in (\ref{mudr})|but now each element $p_i
v^{A}$ has to be repeated $d_A$ times. The indices under the curly
brackets in (\ref{komar1}--\ref{komar3}) indicate now the number of
distinct elements.  Again, we need to divide the vector $\vec{\omega}$ in
(\ref{komar1}--\ref{komar3}) into two equal parts. For simplicity [and 
without altering the asymptotic formulas (\ref{kesh}, \ref{desh})]
assume that $N$ is an odd number. 

If $\alpha$ is an even number, dividing $\vec{\omega}$ 
into two equal parts amounts to finding an integer $m$ 
that satisfies
\BEA
{\sum}_{k=0}^\alpha d_k + {\sum}_{k=1}^m d_{\alpha+k}+{\sum}_{k=0}^m d_k=2^N,
\label{ort}
\EEA
since now the first half of $\vec{\omega}$ ends up with element $p_1v^m$.
This leads to $m=\left\lfloor \frac{N}{2}\right\rfloor -\frac{\alpha}{2}$, where 
$\lfloor x \rfloor$ is defined after (\ref{mudr}).

If $\alpha$ is an odd number, the first half of $\vec{\omega}$ ends up with 
element $p_0v^{\alpha+m}$, and 
$m=\left\lfloor \frac{N}{2}\right\rfloor-\left\lfloor \frac{\alpha}{2}\right\rfloor$ 
is found from 
\BEA
{\sum}_{k=0}^\alpha d_k + {\sum}_{k=1}^m d_{\alpha+k}+{\sum}_{k=0}^{m-1} d_k=2^N.\nonumber
\EEA

\begin{figure}[ht] 
\vspace{0.25cm}
\includegraphics[width=7cm]{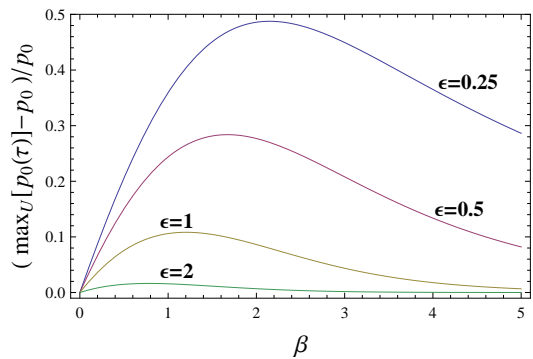} 
\caption{The relative cooling $\frac{{\rm max}_u[p_0(\tau)]-p_0}{p_0}$ calculated
according to (\ref{stuks}, \ref{gaga}) versus the initial inverse temperature $\beta=1/T$ for various values of $\epsilon$ and
$N=201$, $\delta=0.1$. The relative cooling is maximal at some intermediate $\beta(\epsilon)$. The maximum is sharper
for smaller values of $\epsilon$.
} 
\label{f0} 
\end{figure}

For simplicity we focus on the even $\alpha$ case (\ref{ort}).
Then taking in (\ref{komar1}--\ref{komar3}) all the initial 
elements (together with their degeneracies) up to the index 
$m$ means dividing it into two equal parts. Thus the largest 
ground-state probability achievable is equal to the sum of 
the elements from the first half of $\vec{\omega}$:
\BEA
&&{\rm max}_U[\,p_0(\tau)\,]= \frac{1}{(1+v)^N}\left [
p_0 {\sum}_{k=0}^{\alpha+m} \, d_k \, v^k\right.
\nonumber\\
&&+\left.p_1 {\sum}_{k=0}^{m} \,d_k \, v^k
\right], \qquad m=\left\lfloor \frac{N}{2}\right\rfloor-\frac{\alpha}{2}.~
\label{gaga}
\EEA
Examples of (\ref{gaga}) are presented in Fig. \ref{f0}.

Let us now take the limit $N\gg 1$. Despite this limit
both $\epsilon$ and $\delta$ are finite; we neglect 
$\alpha$ in (\ref{gaga}) by taking there
$m=\left\lfloor \frac{N}{2}\right\rfloor$ [recall that $\alpha=0$ for $\epsilon <\delta$, an almost degenerate case] and transform 
(\ref{gaga}) via
\begin{gather} 
\frac{1}{(1+v)^N}{\sum}_{k=0}^{m}d_k v^k = \frac{\sum_{k=0}^{N}d_k v^k - \sum_{k=m+1}^{N}d_k v^k}{(1+v)^N} \nonumber\\
= 1-\frac{v^{m+1} d_{m+1}}{(1+v)^N} {\sum}_{k=0}^{N-m-1} v^k \,\frac{d_{k+m+1}}{d_{m+1}}. 
\label{durer} 
\end{gather}

Since $m=\left\lfloor \frac{N}{2}\right\rfloor>Nv/(1+v)$, $v^k d_{k+m+1}$ is a decaying function of $k$; hence 
${\sum}_{k=0}^{N-m-1} v^k \,\frac{d_{k+m+1}}{d_{m+1}}$ is dominated by its first few terms 
and is bounded from above for $N\to\infty$
\footnote{For numerics one can use the hyper-geometric function:
${\sum}_{k=0}^{N-m-1} v^k
\,\frac{d_{k+m+1}}{d_{m+1}}=\,_2F_1[1,m-N+1,2+m,-v]$ that holds for any $N-m-1>0$. This formula is
derived by expressing $\frac{d_{k+m+1}}{d_{m+1}}$ via Euler's gamma
functions [recall (\ref{sos})], extending the summation to infinity:
${\sum}_{k=0}^{N-m-1}={\sum}_{k=0}^{\infty}$, and employing the standard
definition $_2 F_1 (a,\,b,\,c;\;z)={\sum}_{k=0}^{\infty}\frac{(a)_k
(b)_k}{(c)_k} \cdot \frac{z^k}{k!}$ of the hyper-geometric function
\cite{niki}. Here $(a)_k = \frac{\Gamma(a+k)}{\Gamma(a)}$ and
$(-a)_k=(-1)^k\frac{\Gamma(a+1)}{\Gamma(a-k+1)}$ are the Pochhammer
symbols, and $\Gamma(z)$ denotes the gamma function.  }.
We are left with estimating $\frac{v^{m+1} d_{m+1}}{(1+v)^N}$ for $m=\frac{N}{2}$. 
Using the Stirling's formula (\ref{sti}) 
\BEA
N!\simeq \sqrt{2\pi N}(N/e)^N,
\label{sti}
\EEA
we get for $N\gg 1$
\begin{gather}
\label{kesh}
\frac{1}{N}\ln(1-{\rm max}_U[\,p_0(\tau)\,]\,)=-S[\frac{1}{2}I||\sigma]+{\cal O}(\frac{\ln N}{N}),
\end{gather}
where the relative entropy $S[\frac{1}{2}I||\sigma]$ 
between the probability vectors $(\frac{1}{2},\frac{1}{2})$ and
$(\frac{v}{1+v},\frac{1}{1+v})=\sigma$ satisfies
\begin{gather}
S[\frac{1}{2}I||\sigma]=\ln\frac{1+v}{2\sqrt{v}}=\ln\cosh [\frac{\beta\delta}{2}];
\label{desh}
\end{gather}
see also (\ref{kosa3}). Note that $\sigma$ is the state of the single
reservoir spin [the reservoir consists of $N$ such identical spins].
Note that for $N\gg 1$ the asymptotic behaviour of $p_0(\tau)$ does not
depend on $\epsilon$. 

Thus for $N\to\infty$, the (small) deviation of $p_0(\tau)$ from $1$ is
controlled by the relative entropy (\ref{desh}). The final temperature
of the two-level system reads from (\ref{kesh}, \ref{desh}) 
\BEA
\label{glasgo}
T_{\rm min}=\frac{\epsilon}{NS[\frac{1}{2}I||\sigma]}.
\EEA
Note that (\ref{glasgo}) is consistent with the unattainability argument (\ref{molotov}).

Let us now turn to calculating the minimal work needed for this cooling.

\subsection{The minimal work}

The message of the following calculations is to show that the minimal
work required for cooling (\ref{kesh}, \ref{desh}) {\it can} converge to
its thermodynamic lower bound (\ref{gaspar}) in the thermodynamic limit $N\gg 1$. 

We start by the case $\epsilon=0$. Conceptually, this case of
degenerate target system energy levels is interesting, because no brute
force method (i.e., a low-temperature bath) can ever increase the
occupation of one of the energy levels.  In addition, the work done for
cooling is not blurred by the energy released from the two-level system.
Starting with this case is also useful for technical reasons, moreover
that $\epsilon>0$ will not bring essential news. 

As follows from the discussion around (\ref{ort}), the vector (\ref{komar1}--\ref{komar3}) 
can be written via its two halves as
\BEA
\vec{\omega} = (\vec{a}, \vec{b}),
\EEA
where
\BEA
\label{k1}
&&\vec{a}=\frac{1}{2(1+v)^N}[\, v^0,v^0,\ldots, v^{[N/2]}, v^{[N/2]}],\\
\label{k2}
&&\vec{b}=\frac{1}{2(1+v)^N}[\, v^{[N/2]+1},v^{[N/2]+1},\ldots, v^{N}, v^{N}],\\
&&\vec{c}=[\,0,1,2,\ldots,N\,],
\label{k3}
\EEA
and where $\vec{c}$ is the vector of the reservoir energies. Recall that in 
(\ref{k1}) and (\ref{k2}) each factor $v^{k}$ is repeated $d_k$ times. Likewise, in
(\ref{k3}) each symbol $k$ is repeated $d_k$ times. Vectors
$\vec{a}$, $\vec{b}$ and $\vec{c}$ have equal number of components $2^N$.

The discussion around (\ref{ort}) implies that $\vec{a}$ contains the
largest $2^N$ elements of the vector (\ref{komar1}--\ref{komar3}). Thus
the minimal final reservoir energy needed for the optimal cooling
(\ref{kesh}) is the following inner product
\BEA
\label{art}
\delta(\vec{a}+\vec{b})\vec{c}.
\EEA
Once reservoir's initial energy $\delta Nv/(1+v)$ is known, the
initial and final energies of the target spin are known as well, the minimal work is
determined via (\ref{art}). 

Note that $d_kv^k$ is peaked around 
\BEA
k=\left\lfloor N\frac{v}{1+v} \right\rfloor\equiv N\xi,
\EEA
where $\lfloor x\rfloor$ is defined after (\ref{mudr}), $v=e^{-\beta\delta}$ and 
the difference $\left| \xi - \frac{v}{1+v} \right|$
fluctuates for different $N\gg 1$ only by the amount $1/N$. Hence
in $\vec{a}$ we can focus on dominant energy levels 
\BEA
\label{burnaz}
\{\delta(N\xi +m)\}_{m=-\Delta}^\Delta, 
\EEA
where $\Delta$ is determined by requiring
that the ratio $\frac{d_{N\xi\pm \Delta}v^{N\xi\pm
\Delta}}{d_{N\xi}v^{N\xi}}$ of the central (maximal) probability to the
probability at the edge of the interval (\ref{burnaz}) is exponentially (over $N$)
small. This guarantees that considering only (\ref{burnaz}) suffices for
calculating quantities that are not exponentially small. Employing for
$N\gg 1$ Stirling's formula (\ref{sti}), we get
\BEA
\label{limpopo}
\frac{d_{N\xi\pm \Delta}\,\,v^{N\xi\pm \Delta}}{d_{N\xi}\,\,v^{N\xi}}=
e^{-\frac{\Delta^2}{2\xi(1-\xi)N}+{\cal O}(\frac{\Delta^3}{N^2})+{\cal O}(\frac{\Delta}{N})}.
\EEA
For achieving the sought exponential smallness we need to require
$\Delta=N^{a+1/2}$, where $a$ is a fixed small number, e.g., $a=1/10$.
Hence the total probability of energy levels that do not fall in 
the interval (\ref{burnaz}) is ${\cal O}[Ne^{-N^{2a}}]\to 0$.
Now in (\ref{art}) we can neglect $\vec{b}\vec{c}$ and keep in
$\vec{a}\vec{c}$ only those elements of $\vec{a}$ whose energies lay in the interval (\ref{burnaz}).

Within interval (\ref{burnaz}) we select a segment consisting of
$v^{m+N\xi}$'s only. In $\vec{a}$ it occupies positions with numbers
from $1+\sum_{k=0}^{N\xi+m-1}2d_k$ to $\sum_{k=0}^{N\xi+m}2d_k$.  Its
length is $2d_{m+N\xi}$. It appears that there are only two energies
$\Phi_m-1$ and $\Phi_m$ that correspond to that segment in $\vec{c}$.
We write $\Phi_m=N\xi+m+\ell_m$, where
$\ell_m$ is the minimal integer that satisfies
\BEA
{\sum}_{k=0}^{N\xi+m+\ell_m-1}d_k-{\sum}_{k=0}^{N\xi+m-1}2d_k\equiv D_{m}\geq 0.
\label{gago}
\EEA
The reason of having only two energies for each segment in the dominant interval (\ref{burnaz}) is that 
$\ell_m$ does not depend on $m$, as we see below.

Each sum in (\ref{gago}) is dominated by its largest summand. Recalling from (\ref{sos})
\BEA
\label{aro}
\frac{ d_{k+1} }{ d_{k} }=\frac{1-\frac{k}{N}}{\frac{k}{N}}[1+{\cal O}(\frac{1}{N})],
\EEA
we get a general pattern for approximating such sums
\footnote{Note that both (\ref{aro}) and (\ref{karo}) contain error 
${\cal O}(\frac{1}{N})$. The error ${\cal O}(\frac{1}{N})$ from (\ref{aro}) does not accumulate in 
(\ref{karo}), because the sum ${\sum}_{k=0}^{N\xi}d_k$ is dominated by few (smaller than $N$) terms
around $k=N\xi$.}
\BEA
\label{karo}
{\sum}_{k=0}^{N\xi}d_k =\frac{d_{N\xi}}{1-v}[1+{\cal O}(\frac{1}{N})].
\EEA
Eq.~(\ref{gago}) re-writes as [neglecting factors ${\cal O}(\frac{1}{N})$]
\begin{gather}
{\sum}_{k=N\xi+m+1}^{N\xi+m+\ell_m}\,\,d_k=d_{N\xi+m}{\sum}_{k=1}^{\ell_m}\frac{1}{v^k} \nonumber\\
\geq {\sum}_{k=0}^{N\xi+m}\,d_k=\frac{d_{N\xi+m}}{1-v},
\label{paton}
\end{gather}
These relations imply
\BEA
\label{le}
\ell_m=\ell=\left\lceil\frac{\ln 2}{\ln\frac{1}{v}}\right\rceil=\left\lceil\frac{\ln 2}{\beta\delta}  \right\rceil,
\EEA
where $\lceil x\rceil$ is the ceiling integer part of $x$, e.g. $\lceil 0.99\rceil=1$, $\lceil 1.1\rceil=2$.
As anticipated, $\ell_m$ does not depend on $m$ \footnote{This fact together with (\ref{le}) and the
reasoning above it allows to guess an upper bound $\delta\ell$ for the work $W$. This bound is confirmed by  
(\ref{of}) and allows to deduce quickly the fact of (\ref{khanum}).}. Hence we write for the minimal work (\ref{art})
\begin{gather}
\delta\vec{a}\vec{c}=
\frac{\delta}{2(1+v)^N}{\sum}_{m=-\Delta}^{\Delta}\,v^{N\xi+m} [ D_m(N\xi+m+\ell-1)\nonumber \\
+(2d_{N\xi+m}-D_m)(N\xi+m+\ell)],
\label{ashun}
\end{gather}
where $D_m$ is defined in (\ref{gago}). Eq.~(\ref{paton}) implies:
\BEA
D_m= d_{N\xi+m}\frac{v(v^{-\ell}-2)}{1-v}[1+{\cal O}(\frac{1}{N})].
\label{turbo}
\EEA
Note that
$2d_{N\xi+m}-D_m>0$ 
for the considered range of $m$ \footnote{In view of (\ref{turbo}) this reduces to $\frac{2}{v}>\frac{v^{-\ell}-2}{1-v}$.}. 
Using (\ref{turbo}) we get for (\ref{ashun}) 
\BEA
\delta{\sum}_{m=-\Delta}^{\Delta}\,\frac{d_{N\xi+m}v^{N\xi+m} }{(1+v)^N}
[N\xi+\ell-\frac{v(v^{-\ell}-2)}{2(1-v)}+m]\nonumber
\EEA
Neglecting \footnote{Recall (\ref{sos}) and denote $\widetilde{d}_A=\frac{(N-1)!}{(N-A-1)!\,A!}$. 
Put $m=N\xi+m-N\xi$ in ${\sum}_{m=-\Delta}^{\Delta}\,\frac{m d_{N\xi+m}v^{N\xi+m} }{(1+v)^N}$
and using $kd_k=N\widetilde{d}_{k-1}$ write it as
$$\xi N\{ {\sum}_{m=-\Delta-\widetilde{\xi}}^{\Delta-\widetilde{\xi}}\,\,
\frac{\widetilde{d}_{\widetilde{N}\xi+m}v^{\widetilde{N}\xi+m} }{(1+v)^{\widetilde{N}}}
-{\sum}_{m=-\Delta}^{\Delta}\,\frac{ d_{N\xi+m}v^{N\xi+m} }{(1+v)^N}\},$$ where $\widetilde{N}=N-1$ and $\widetilde{\xi}=1-\xi$. 
Each sum in the curly brackets is equal to $1$ minus exponentially small terms; cf with (\ref{limpopo}).
} ${\sum}_{m=-\Delta}^{\Delta}\,\frac{m d_{N\xi+m}v^{N\xi+m} }{(1+v)^N}$
we get for the work
\BEA
W=\delta\left [\ell-\frac{v(v^{-\ell}-2)}{2(1-v)}
\right].
\label{of}
\EEA
This expression implies several important conclusions:

$\bullet$ The initial and final reservoir energy differ from each other
by a factor $W={\cal O}(1)$; hence the state of each reservoir particle
changes by a quantity of order ${\cal O}(1/N)$, neglegible for $N\gg 1$. 

$\bullet$
When $\frac{\ln 2}{\beta\delta}$ approaches to an integer number from
below, (\ref{le}, \ref{of}) predict: 
\BEA
W\to T \ln 2.
\label{khanum}
\EEA

\begin{figure}
\includegraphics[width=7.5cm]{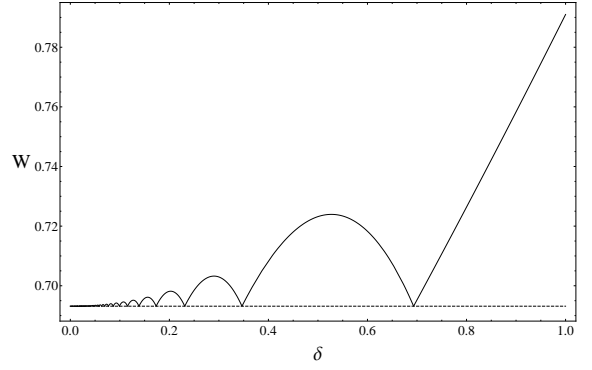}
\caption{The minimal work $W(\delta)$ for $\epsilon=0$ and $T=1$. The dotted line is $T\ln 2$, 
the lower bound (\ref{gaspar}) for $W(\delta)$. }
\label{kapur}
\end{figure}

According to (\ref{gaspar}) [and recalling that $\epsilon=0$] $T\ln 2$
is the minimal possible work necessary to cool from the absolutely
disordered state [the equilibrium state with $\epsilon=0$] to an almost
ordered state (\ref{kesh}).  Hence the thermodynamic bound
(\ref{gaspar}) is reachable for certain {\it finite} $\delta$'s. The
behavior of $W(\delta)$ is shown in Fig.~\ref{kapur}. For a finite $N$,
$W$ is larger than $T\ln 2$ even if $\frac{\ln 2}{\beta\delta}$ is an
integer; see Table \ref{tab0} for numerical results. 

\begin{table}
\caption{
The minimal work $W$ as a function of $N$ for $\delta=\ln 2=0.693147$, $\epsilon=0$ and $T=1$.
}
\begin{tabular}{|c||c|}
\hline
\,\,$N$  \,\,           & ~$W$~   \\
\hline \hline
\,\, $N=601$   \,\,    & \,0.716891\,  \\
\hline
\,\, $N=1001$   \,\, & \,0.711592\,    \\
\hline
\,\, $N=1401$   \,\, & \,0.708766\,   \\
\hline
\,\, $N=2001$   \,\, & \,0.70623\,  \\
\hline
\end{tabular}
\label{tab0}
\end{table}

$\bullet$ For $\delta>T\ln 2$, (\ref{of}) yields
$W(\delta)=\frac{\delta}{2(1-v)}$: now the minimal
work monotonically increases with $\delta$ (reservoir energy gap).
Recall from (\ref{kesh}, \ref{desh}) that for a large but finite $N$ the
deviation of ${\rm max}_U[\,p_0(\tau)\,]$ from $1$ is controlled by
$S[\frac{1}{2}I||\sigma]$, which also grows with $\delta$. Thus enhanced
cooling demands more work. Moreover, one observes:
\BEA
W\approx
\Delta F + TS[\rho||\sigma], ~~{\rm for}~~ T\ln 2\leq \delta,
\label{gogen}
\EEA
where $\Delta F=T\ln 2$ is the free-energy
difference from (\ref{gaspar}), $\rho=\frac{1}{2}I$ [for $\epsilon=0$]
is the initial state of the target two-level system, and $\sigma$ is the
initial state of any reservoir spin. Eq.~(\ref{gogen}) becomes exact for $\delta\gg T\ln 2$. 

The case $\epsilon>0$ does not require any new idea as compared to the above derivation; only algebraic 
steps are more tedious. Hence we quote only the final result for the minimal work:
\BEA
W=\left\{ \begin{array}{l} \delta\lb \ell -\frac{f-u\al}{1+u} \rb -\frac{\epsilon u}{1+u}; 
~{\rm for}~ f\leq 1 \\ \\ \delta\lb \ell -\frac{1+u v^{-\al}(f-1)-u\al}{1+u} \rb -\frac{\epsilon u}{1+u}; 
~{\rm for}~ f> 1 \end{array} \right.
\label{kaban}
\EEA
where $v=e^{-\beta\delta}$, $u=e^{-\beta\epsilon}$, 
$\alpha$ is defined in (\ref{mudr}), and where
\BEA
\ell=\left\lceil\frac{\ln (1+v^\alpha)}{\ln\frac{1}{v}}\right\rceil, ~~
f=\frac{v(v^{-\ell}-1-v^\al)}{1-v}.
\EEA
For $\epsilon=0$ both branches of (\ref{kaban}) are equal, so no conditioning is needed. 

Our conclusion on the reachability of the thermodynamic lower bound
(\ref{gaspar}) remains valid also for $\epsilon>0$: consider $\frac{\ln
(1+v^\alpha)}{\ln\frac{1}{v}}$ reaching an integer number from below,
and $\frac{\epsilon}{\delta}$ reaching an integer number from above.
Then (\ref{kaban})
produces $W=T\ln(1+e^{-\beta\epsilon})$. This is the bound (\ref{gaspar})
[recall (\ref{kesh}, \ref{desh})].

\subsection{Microcanonical initial state of the reservoir}
\label{micro}

So far we considered the reservoir starting its evolution from a
canonical equilibrium state (\ref{mu}, \ref{brams}). Another notion of
equilibrium is provided by the microcanonical density matrix, where all
the energy levels in the vicinity $[E-\kappa,E+\kappa]$ of a
given energy $E$ have equal probability, while all other energy levels
have zero occupation. While the canonical state describes a system
prepared via a thermal bath, the microcanonical state refers to a closed
macroscopic system whose energy is fixed up to a small uncertainty
$2\kappa$ \cite{q_t,landau}. Hence $\kappa$ should be large
enough for $[E-\kappa,E+\kappa]$ to incorporate many energy
levels and account for unavoidable environmental perturbations, but
small compared with $E$ \cite{q_t,landau}. 

The canonical and microcanonical notions of equilibria are normally
equivalent for macroscopic observables of systems containing
sufficiently many particles; see \cite{touchette} for a review. The
microcanonical state can apply for systems with a finite number
particles provided that the interaction Hamiltonian does have certain
chaoticity features; see \cite{q_t,borgonovi} and refs. therein. There
are situations|such as systems in the vicinity of phase transitions, or
systems with long-range interactions|where the equivalence between
canonical and microcanonical state is broken even for macroscopic
observables; see, e.g., \cite{touchette,kastner,leyvraz} and
\cite{campa} for review. Generally, these states are different regarding
some important aspects \cite{kastner,leyvraz,campa}. In particular, the
local stability conditions (i.e., stability iwht respect to small
perturbations) for the microcanonic state are less restrictive than for
the canonic state (with the same temeprature) \cite{campa}. Hence it
happens in several physically interesting situations that the canonic
state is unstable, and the microcanonic state is the only description of
the equilibrium \cite{kastner,leyvraz,campa}.

For the reservoir consisting of $N\gg 1$ identical spins the
microcanonical state concentrated at the energy $\delta K$ with $K\gg 1$
is especially easy to define: the $d_K$ degenerate energy levels
$\delta K$ [see (\ref{sos})] have equal probability $\frac{1}{d_k}$, 
while all other energies have zero probability. Because the energy level
$\delta K$ is already exponentially degenerate we restricted ourselves
with the minimal width $\kappa\to 0$. 

Given this initial state of the reservoir it is easy to see using the
same construction as in (\ref{komar1}--\ref{komar3}) [see our discussion
after (\ref{sos})] that the maximal final ground state probability for
the two-level interacting with such a reservoir is just equal to $1$:
\BEA
{\rm max}_{U}[\,p_0(\tau)\,]=1.
\label{lao}
\EEA
The result (\ref{lao}) does not require $N\to \infty$ or $\kappa\to 0$. It is also not
specific to the microcanonical density matrix. Any density matrix for an
$M$-level reservoir that is diagonal in the energy representation, and
that has at least $\left\lfloor\frac{M}{2}\right\rfloor$ zero eigenvalues would lead to
(\ref{lao}); cf. with our discussion in section \ref{nozero}. For
example, it also applies to the so called $\theta$-ensemble, where all
the energy levels below a certain $E$ are equally populated, while the energies
above $E$ have zero probability. 

Eq.~(\ref{lao}) was obtained via the standard definition of the
microcanonical state, but it contains an essential idealization: it is
assumed that at least $\left\lfloor\frac{M}{2}\right\rfloor$ energies have exactly zero
probability. Consider the emergence of the microcanonic state.
Classically, any single system is in a state with definite energy. Weak
and inevitable interactions with environment smear this energy over the
interval $2\kappa$.  This however does not ensure that energies
outside $[E-\kappa,E+\kappa]$ have strictly zero probability.

Quantum mechanically, we cannot know the state of a single system unless
we prepared it, e.g., by measurement. One [the only?] general method of
preparing a microcanonical state is to do a selective measurement of
energy and then isolate the system \footnote{The selection will produce
(Luders' postulate) a density matrix $\propto\Pi(E,\kappa)\rho_{\rm
in}\Pi(E,\kappa)$, where $\rho_{\rm in}$ is the initial state of the
system, and where $\Pi(E,\kappa)$ is the projector on the Hilbert
subspace with energies $[E-\kappa, E+\kappa]$. If $\rho_{\rm in}$ was a
sufficiently smooth function of the Hamiltonian (e.g., a Gibbs state), a
microcanonic state results. If $\rho_{\rm in}$ was an arbitrary state we
face an additional problem of relaxation towards the microcanonic
state.}. Generally, after a selective measurement all energies will be
populated, because the measurement itself may be noisy due to our
inability to control the measurement interaction and/or the system
Hamiltonian. Alternatively, the system may be weakly coupled to its
environment prior to the energy measurement, which then necessarily
refers to measuring the local energy value. For an open system the
latter observable is not strictly conserved \footnote{If the
environmental coupling is not weak, the local energy has to be properly
defined; see \cite{q_t} for a concrete proposal (LEMBAS principle).}.
Again, both these mechanisms will generally populate energies outside
the interval $[E-\kappa,E+\kappa]$.  It now suffices to give a small,
but non-zero probability to more than
$\left\lfloor\frac{M}{2}\right\rfloor$ energy levels, and ${\rm
max}_{U}[\,p_0(\tau)\,]$ in (\ref{lao}) will be smaller than one, as we
saw in section \ref{nozero}. 

Hence, the unattainability of $T=0$ for the microcanonic state of the
reservoir is recovered provided the microcanonic state contains such
tails.  Admittedly, we currently lack any general condition for how
small these tails of the microcanonic state could be made. It is also
unclear which {\it generic} [not {\it ad hoc}] conditions one has to
impose on the coupling to environment or on the measurement noise to
ensure the tails needed for unattainability. In this context one
anticipates an additional tractability condition {\bf T4}|referring to a
finite environmental coupling or a finite measurement noise|to be added
to conditions {\bf T1--T3} in section \ref{summa_contra} and
\ref{comparo} \footnote{Conversely, if the unattainability of the
absolute zero is regarded to be a law of nature, it can constrain the
microscopic processes of environmental coupling and selective measurement. Albeit
this possibility contradicts to the current paradigm of deriving the
laws of thermodynamics from microscopic theories, it needs to be taken
seriously.}.

Note the difference: the reachability of low temperatures for a
canonical reservoir with a fixed upper energy would require an ever
increasing energy gap (related to an ever increasing work cost), while 
for a microcanonic reservoir (with a fixed upper energy again) 
reaching low temperatures requires small tails around the central energy. 

\begin{figure}
\includegraphics[width=7.5cm]{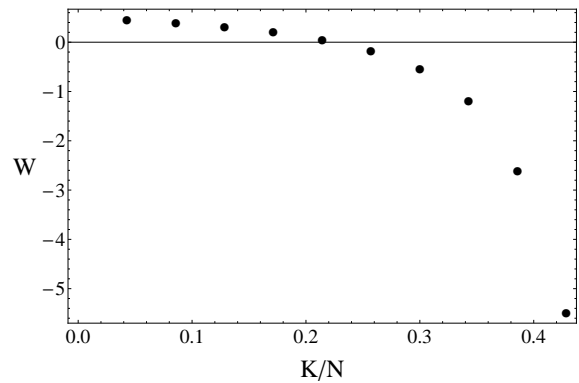}
\vspace{0.5cm}
\caption{
The minimal work $W$ (needed for maximal cooling) versus the initial energy $\delta\,K$ of the microcanonical state,
where the initial state of the $N$-spin reservoir is the mixture of
$\frac{N!}{K! (N-K)!}$ energy eigen-states with the energy $\delta
K$. We considered: $N=7\times 10^4$, $\delta=1$, $\epsilon =0$. 
}
\label{f5}
\end{figure}

Another essential difference is that there is no work cost associated
with the microcanonic reservoir: the minimal work necessary for the
maximal cooling need not be positive (i.e. no work has to be consumed);
see Fig.~\ref{f5}. This is related to the general fact that the
microcanonical ensemble is not passive \cite{passivity}: there exist for
it a class of unitary operations [generated by a suitable cyclically
changing Hamiltonians] that leads to work-extraction \cite{passo}.
Importantly, the unitary realizing the maximal cooling for the
microcanonical reservoir can be put into that class, as Fig.~\ref{f5}
shows.

\section{Summary}

In this paper we have studied dynamical cooling of a two-level system
(as target) in contact to various types of quantum reservoirs prepared
in various types of initial states. Based on operational ideas these
schemes have been placed within the context of given resource
constraints; the operational meaning of the third law can thus be
clarified. 

\subsection{Canonical initial state. Complementarity betweeen cooling and work}

Unattainability of $T=0$ is a direct consequence of the dynamic cooling
set-up; see section \ref{nozero}. In contrast to the unattainability
formulation of the third law, it does not involve unproven or disguised
assumptions. For a given reservoir with a fixed upper energy starting
its evolution from a canonical equilibrium, and for a two-level system
as the target of cooling, we explicitly predict the lowest non-zero
temperature (\ref{dardibun}) achievable within dynamic cooling. The lowest
temperature (\ref{dardibun}) would approach zero whenever the upper
energy of the reservoir goes to infinity. 

We have been able to re-interpret this functional dependence in terms of
the work intake: reaching the lowest temperature demands work growing as
the inverse of this temperature; see (\ref{trdao}). However, this linear
relation only applies for reaching that nominal lowest temperature
possible: compromising for a slightly increased final target temperature
would allow to change the work growth regime from the linear to
(multiple) logarithmic; see section \ref{trade-off}.

Next, we studied concrete models of reservoir. For them we found the
minimal temperature attainable (and the minimal work necessary to reach
it) by optimizing over cooling dynamics. In section \ref{N} we modeled
the reservoir via a thermal bath consisting of a large number, $N\gg 1$,
of identical spins, and established relations between the resources of
cooling and certain relative entropies. The lowest temperature $T_{\rm
min}$ reachable with such a reservoir scales as the inverse of $N$ and
the relative entropy $S[\frac{1}{2}I||\sigma]$, where $I$ is the
$2\times 2$ unit matrix and $\sigma$ is the initial state of a bath
spin; see (\ref{glasgo}). The minimal work $W$ needed to reach this
temperature has basically two different expressions. If the reservoir
gap is sufficiently larger than temperature, $\delta\gtrsim T$, we get
$W\to \Delta F +S[\rho ||\sigma]$, where $\Delta F$ is the free energy
difference of the cooled two-level system, and where $\rho$ is its
initial (hot) state. Recall that $\Delta F$ is the minimal work required
by thermodynamics for achieving cooling; see (\ref{gaspar}) and
(\ref{a3}).  If $\delta\lesssim T\ln 2$, $W$ is close to $\Delta F=T\ln
2$.  Hence enhancing cooling demands a larger work, but a substantial
cooling can already be achieved with the minimal work $\Delta F$
demanded by the second law. We should thus stress that approaching
$T=0$, while using only the amount of work $\Delta F$, is feasible for
several limits of reservoir parameters. 

Note that the existence of the lowest temperature is related to a
finite upper energy of the reservoir (either it really has this feature,
or one considers only that class of cooling operations, where only
finite reservoir energies couple to the target of cooling). If the upper
energy of the reservoir is infinite, the unattainability of $T=0$ is
still there, but the minimum temperature does not exist. Instead, the
infimum temperature is zero; see section \ref{horror}.

\subsection{Microcanonical initial state}

The unattainability of $T=0$ can be recovered also for a reservoir
starting its evolution from a microcanonical state; see section
\ref{micro}. However, the origin of the strictly non-zero minimal
temperature is different: It relates to tails of the energy distribution
resulting from a weak interaction with the environment (and thus our
inability to avoid such marginal couplings). Put differently, the
difference between canonical and microcanonical reservoir concerns the
physics of the $T\to 0$ limit. Low temperatures in the canonical
situation require a large energy gap in the reservoir. 
For the microcanonical case low temperatures require a sufficiently
large system which can be prepared in a state with an almost definite
energy. There is no work cost involved. 

\subsection{Comparison with other approaches}

Ref.~\cite{Dom_Beth} studies a cooling set-up with three components: the
system ${\bf S}$ to be cooled, the resource ${\bf R}$, and the
environment ${\bf E}$ with Hamiltonians $H_{\bf S}$, $H_{\bf R}$, and
$H_{\bf E}$, respectively.  {\bf S} and {\bf E} start in, respectively,
the Gibbs states $\gamma_{\bf S}$ and $\gamma_{\bf E}$ with the same fixed
initial temperature $T_{\rm in}$. Then, \cite{Dom_Beth} allows only for
cooling operations that are unitary on the Hilbert space of ${\bf
S+R+E}$ and commute with the interaction free joint Hamiltonian $H_{\bf
S}+H_{\bf R}+H_{\bf E}$.  Hence, no work cost is related to implementing
this unitary and cooling is only possible if the resource deviates from
its Gibbs state $\gamma_{\bf R}$ at the same temperature $T_{\rm in}$, i.e., a
resource being in its Gibbs state is "worthless"; see in this context
(\ref{mega}--\ref{gaspar}).  If the resource ${\bf R}_N:={\bf R}^{\times
N}$ consists of $N$ identical copies, the lowest temperature that can be
achieved for ${\bf S}$ is asymptotically determined by an analogue of
(\ref{glasgo}), where instead of $S[\frac{1}{2}I||\sigma]$ one has
$S[\gamma_{\bf R} ||\rho_{\bf R}]$ \cite{Dom_Beth}. Here $\gamma_{\bf
R}$ is the Gibbs state and $\rho_{\bf R}$ the actual state of one
resource copy. This relation between (\ref{glasgo}) and the results of
\cite{Dom_Beth} relates to the fact that in our present setting the
reservoir that consists of $N$ identical subsystems can (asymptotically)
be used for cooling if and only if the subsystems are not in their
maximally mixed state $\frac{1}{2}I$; see section \ref{N}.  Hence,
relative entropies other than free energies determine the value of a
resource (reservoir) both in the present paper and in \cite{Dom_Beth}.
Note that the relative entropy $S[\gamma_{\bf R} ||\rho_{\bf R}]$ must
not to be confused with $S[\rho_{\bf R} || \gamma_{\bf R}]$, which is
essentially the free energy up to constants. As opposed to free energy,
$S[\gamma_{\bf R} ||\rho_{\bf R}]$ diverges if $\rho_{\bf R}$ gets a
zero eigenvalue. Thus the absolute zero is reachable only for a
resource state having a zero eigenvalue, an aspect closely related to
the microcanonic treatment of section \ref{micro}.

Alternatively, the features of the limit $T\to 0$ can be studied via
refrigerators that cyclically operate between two thermal baths at
temperatures $T_c$ and $T_h$ ($T_c<T_h$) and cool the (finite)
low-temperature bath at the expense of consuming work from an external
source \cite{yan,gordon,rezek,segal,karen}. If the refrigerator works at
a finite efficiency, then for $T_c\to 0$ the heat taken per unit of time
from the low-temperature bath scales as $T_c^{a+1}$ with $a\geq 0$
\cite{gordon,rezek}, showing that cooling the low-temperature bath is
progressively slowed down. The optimal behavior $a=0$ is reached for the
refrigerator model studied in \cite{karen}. 

\subsection{Open issues}

An important aspect of practical implications is the issue of time
required to complete the cooling transformation. In our present
operational analysis we have dealt with this problem rather formally, assuming
that once the needed unitary transformation is constructed, there is
always a (time-dependent) Hamiltonian that realizes it. Provided that we
do not restrict the magnitude of the external fields realizing the
Hamiltonian, this realization could, in principle, take an arbitrary
short time, i.e., the external fields can be assumed to function in the
pulsed regime. However, if we demand that the optimal cooling unitary
transformations are implemented in terms of a well-defined base of {\it
realizable} unitary operations, the cooling process may turn out
to be complex: it may take a long sequence of the realizable operations
to contruct the needed unitary.  This complexity (and hence time)
resource needs further studies. 

Another open issue concerns target systems with more than two energy
levels.  This should be interesting especially with respect to
non-equilibrium aspects. Moreover, cooling to very low temperatures may
in fact demand considering many-body targets of cooling, because at such
low temperatures the standard assumptions of the weak coupling between
the target and the reservoir|as well as between different parts of the
target|may be broken \cite{theo}. We nevertheless expect that the basic
features of dynamic cooling discussed here will survive for more general
models (the unattainability of the ground state and a non-trivial work
cost for cooling).

\subsection*{Acknowledgement}
This work has been supported by Volkswagenstiftung.

We thank R. Kosloff for discussions.

\appendix

\section{}
\label{apo}

Here we show that the unitaries implementing optimal cooling can be
chosen to be permutations of energy eigenstates. 

We are given two sequences $\Lambda = (\lambda_1, . . . ,\lambda_N)$ and
$D= (d_1, . . . ,d_N)$ of real numbers. Let
$\{\lambda^{\downarrow}_k\}_{k=1}^N$ and $\{d^{\downarrow}_k\}_{k=1}^N$
be the non-increasing arrangements of their elements:
\BEA
\lambda^{\downarrow}_1\geq \lambda^{\downarrow}_2\ldots \geq\lambda^{\downarrow}_n, \quad 
d^{\downarrow}_1\geq d^{\downarrow}_2\ldots \geq d^{\downarrow}_n.
\EEA
$\Lambda$ {\it majorizes} $D$ if the following $N$ conditions hold \cite{olkin}:
\BEA
&&{\sum}_{k=1}^m \lambda^{\downarrow}_k \geq {\sum}_{k=1}^m d^{\downarrow}_k ~~ {\rm for}~~m=1,\ldots,N-1,~~ \\
&&{\sum}_{k=1}^m \lambda^{\downarrow}_k = {\sum}_{k=1}^m d^{\downarrow}_k.
\EEA
In words: for each $m = 1,\ldots ,N$, the sum of the $m$ largest elements of $\Lambda$ is at least as
large as the sum of the $k$ largest elements of $D$, with equality for $m = N$.

Birkhoff showed that $\Lambda$ majorizes $D$ if and only if there is
a double-stochastic matrix $S$ such that 
\BEA
\label{doremi}
D = S\Lambda, 
\EEA
where double-stochastic means that $S_{ij}\geq 0$, $\sum_{j=1}^N S_{ij}=\sum_{i=1}^N S_{ij}=1$ \cite{olkin}. 

Let $\Lambda$ be the sequence (\ref{in}) of initial eigenvalues. Denote
by $\{|i\rangle\}_{i=0}^{2M-1}$ the set of eigenvectors for the initial
Hamiltonian (\ref{brams}). Let $D=\{\langle i|\rho_{\bf
S+R}(\tau)|i\rangle\}_{i=0}^{2M-1}$ be the probability vector of energy
level occupations in the final state (\ref{fio}). Eq.~(\ref{fio})
implies (\ref{doremi}) with a double-stochastic matrix $S_{ij}=| \langle
i|U|j\rangle|^2$, where the unitary operator $U$ is defined by
(\ref{ga}, \ref{fio}). 

The Birkhof theorem implies that the sum of the largest $k$ elements of
$D$ is not larger than that of $\Lambda$. For the purpose of cooling we
want to make the sum of $M$ largest elements of $D$ as big as possible,
and thus it has to be equal to the sum of $M$ largest elements of
$\Lambda$. This is realized if $S$ {\it permutes} the $M$ largest
elements of $\Lambda$. 

\section{}
\label{apa}

The example (\ref{fi}) was developed for $M=2n=4$. It can be generalized to arbitrary $n$:
\begin{gather}
\label{crimp}
\hat{p}_0=\pi_0+\pi_1+\ldots+\pi_{n-1},\qquad k=0,\ldots,n-1,\\
\hat{\pi}_{2k}=p_0(\pi_{k}+\pi_{n+k}),\quad
\hat{\pi}_{2k+1}=p_1(\pi_{k}+\pi_{n+k}),\\
W=\epsilon (\hat{p}_1-p_1) + {\sum}_{k=1}^{2n-1}\mu_k (\hat{\pi}_k-\pi_k).
\end{gather}
We assume 
\BEA
\label{gogo}
\beta\mu_k\equiv \beta\mu\gg 1 \qquad {\rm for}\qquad  k\geq n. 
\EEA
Hence
\begin{gather}
\pi_{k\geq n}\to 0, \qquad \hat{p}_0 \to 1,\\
W=-\epsilon p_1 + {\sum}_{k=1}^{n-1}\mu_k (\hat{\pi}_k-\pi_k)+ {\sum}_{k=n}^{2n-1}\mu_k \hat{\pi}_k\\
 =-\epsilon p_1 + {\sum}_{k=1}^{n-1}\mu_k (\hat{\pi}_k-\pi_k)+ \mu_{n}{\sum}_{k=n}^{2n-1}\hat{\pi}_k.
\end{gather}
Consider in detail the case $n=4$, i.e., $8$-level reservoir:
\BEA
\beta W &=&p_1\ln\frac{p_1}{1-p_1}+\beta \mu_1(p_1\pi_0-\pi_1)\nonumber\\
&+&\beta \mu_2(p_0\pi_1-\pi_2)+\beta \mu_3(p_1\pi_1-\pi_3)+\beta\mu_4(\pi_2+\pi_3)\nonumber\\
&=&\left[1+e^{-\beta\mu_1}+e^{-\beta\mu_2}+e^{-\beta\mu_3}\right]^{-1}
\,[\,\beta \mu_1(p_1-e^{-\beta\mu_1})\nonumber\\
&+&\beta \mu_2(\,p_0 e^{-\beta\mu_1}-e^{-\beta\mu_2})+\beta \mu_3(\,p_1e^{-\beta\mu_1}-e^{-\beta\mu_3})\nonumber\\
&+& \beta\mu_4(e^{-\beta\mu_2}+e^{-\beta\mu_3})  \,],
\label{kobo}
\EEA
where we used (\ref{stuks}, \ref{lenin}).
We now minimize this expression over $\mu_1\leq \mu_2\leq \mu_3$ 
assuming that $\beta\mu_4\gg 1$ is fixed [see (\ref{gogo})].
First we introduce new variables $y_2$ and $y_3$,
\BEA
\label{mro}
\beta\mu_2=\ln\left[\frac{\beta\mu_4}{y_2}\right], \quad
\beta\mu_3=\ln\left[\frac{\beta\mu_4}{y_3}\right], 
\EEA
substitue them into (\ref{kobo}) and obtain after minimization:
\BEA
y_2 = p_0 e^{-\beta\mu_1}, \qquad y_3 = p_1 e^{-\beta\mu_1}.
\label{gel}
\EEA
Note that $y_2>y_3$ seen from (\ref{gel}) due to $p_0>p_1$ is consistent with
$\mu_2<\mu_3$. Next, we put (\ref{gel}) back into (\ref{kobo}):
\BEA
\beta W &=&p_1\ln\frac{p_1}{1-p_1} \nonumber\\
&+&\frac{\beta\mu_1 p_1+e^{-\beta\mu_1}\ln[\beta\mu_4 e^{1+h[p_0]}]}{1+e^{-\beta\mu_1}},
\label{berkut} 
\EEA
where $h[p]\equiv -p\ln p-(1-p)\ln (1-p)$.
Now the RHS of (\ref{berkut}) is to be minimized over $\mu_1$. This is done similarly to 
(\ref{blo}--\ref{mlo}). Provided that $\ln[\beta\mu_4 e^{1+h[p_0]}]$ is sufficiently large, 
the maximization over $\beta\mu_1$ produced
\BEA
\label{gane}
\beta\mu_1 = \ln\left[\frac{1}{p_1}\ln[\beta\mu_4 e^{1+h[p_0]}] \right],\\
\beta W =p_1\ln\frac{p_1}{1-p_1}+p_1\ln\left[\frac{e}{p_1}\ln[\beta\mu_4 e^{1+h[p_0]}] \right].
\EEA

Hence the work needed for cooling scales double-logarithmically with maximal energy gap $\mu_4$. The deviation of 
the ground-state probability (\ref{crimp}) from its maximal value (\ref{dardibun}) is controlled by (\ref{goris});
see (\ref{mro}, \ref{gel}, \ref{gane}) in this context. 

Eqs.~(\ref{mro}, \ref{gane}) show that the reservoir spectrum is
self-similar: $\beta\mu_3$ and $\beta\mu_2$ depend logarithmically on
$\beta\mu_4$, while $\mu_1$ depend on $\beta\mu_4$
doubly-logarithmically. 

Continuing this was one can show for a $2^{m-1}$ level reservoir
the work can scale $\ln[\ldots [\ln [\beta\mu]]\ldots ]$ (logarithm is repeated $m$ times) as a function of the gap.


\begin{thebibliography}{10}

\bibitem{callen} H.B. Callen, {\it Thermodynamics} (John Wiley, NY, 1985). 

\bibitem{wheeler} J. C. Wheeler, Phys. Rev. A {\bf 43}, 5289 (1991); {\it ibid.} {\bf 45}, 2637 (1992).

\bibitem{landsberg} P. Landsberg, Am. J. Phys. {\bf 65}, 269 (1997).

\bibitem{grif}R. B. Griffiths, J. Math. Phys. {\bf 6}, 1447 (1965); 
in {\it A Critical Review of Thermodynamics},
ed. by E. B. Stuart, B. Gal-Or and A. J. Mainard (Mono Book Corp., Baltimore, 1970).

\bibitem{klein}M. J. Klein, in {\it Thermodynamics of Irreversible Processes, 
Scuola internazionale di fisica "Enrico Fermi"}, ed. by S. R. de Groot (Bologna, 1960).

\bibitem{callen_note} H.B. Callen, in {\it Modern Developments in Thermodynamics}, ed. by B. Gal-Or (J. Wiley \& Sons, NY, 1974).

\bibitem{falk} G. Falk, Phys. Rev. {\bf 115}, 249 (1959). 
\comment{
Third Law of Thermodynamics

Gottfried Falk, Phys. Rev. 115, 249 (1959) 

A new formulation of the third law is proposed stating a universal
connection between the lower limits of the energy and the entropy of any
physical system. As consequences of the new theorem are derived the
Nernst heat theorem, a theorem concerning the lowest energy state of
mixtures, and the nondegeneracy of the energetic ground state of
physical systems. }

\bibitem{magnetocaloric}
K. A. Gschneidner Jr, V. K. Pecharsky and A. O. Tsokol, Rep. Prog. Phys. {\bf 68}, 1479 (2005).

\bibitem{abo} A. Abragam and M. Goldman, Rep. Prog. Phys., {\bf 41}, 395 (1978). 

\bibitem{ernst}
R.R. Ernst, G. Bodenhausen, A. Wokaun, {\it Principles
of Nuclear Magnetic Resonance in One and Two Dimensions}
(Oxford University Press, 1987).


\bibitem{ole}O.W. Sorensen, Prog. NMR Spectr., {\bf 21}, 503 (1989).

\bibitem{slichter} C.P. Slichter, {\it Principles of Magnetic 
Resonance} (Springer, Berlin, 1990).

\bibitem{suter}D. Suter, J. Chem. Phys. {\bf 128}, 052206 (2008).

\bibitem{exp}
D.A. Hall, {\it et al.}, Science, {\bf 276}, 930 (1997).
J.H. Ardenkj\ae r-Larsen, {\it et al.}, {\it PNAS}, {\bf 100}, 10158 (2003).

\bibitem{algol} P. O. Boykin {\it et al.}, Proc. Natl. Acad. Sci. U.S.A. {\bf 99}, 3388
(2002).

J. M. Fernandez {\it et al.}, Int. J. Quantum. Inform. {\bf 2}, 461 (2004).

F. Rempp {\it et al.}, Phys. Rev. A 76, 032325 (2007).

\bibitem{schulman} L. J. Schulman {\it et al.}, PRL {\bf 94}, 120501 (2005).

\bibitem{a} A.E. Allahverdyan {\it et al.}, Phys. Rev. Lett. {\bf 93}, 260404 (2004).

\bibitem{lasercooling} J. Eschner {\it et al.}, J. Opt. Soc. Am. B, {\bf
20}, 1003 (2003).

\bibitem{nocooling} W. Ketterle and D.E. Pritchard, Phys. Rev. A {\bf 46}, 4051 (1992).

A. Bartana, R. Kosloff and D. J. Tannor, J. Chem. Phys. {\bf 106}, 1435
(1997); {\it ibid} {\bf 99}, 196 (1993). 

\bibitem{bri} P.W. Bridgman, {\it The Nature of Thermodynamics} (Harvard University Press, Cambridge, 1941).

\bibitem{Dom_Beth}
D. Janzing, P. Wocjan, R. Zeier, R. Geiss and Th. Beth, Int. Jour. Theor. Phys. {\bf 39}, 2217 (2000).
%

\bibitem{derrida} B. Derrida, Phys. Rev. B {\bf 24}, 2613 (1981).

\bibitem{yan} Z. Yan and J. Chen, J. Phys. D {\bf 23}, 136 (1990).

S. Velasco, J. M. M. Roco, A. Medina, and A. C. Hernandez,
Phys. Rev. Lett. {\bf 78}, 3241 (1997).

\bibitem{gordon}R. Kosloff, E. Geva, and J. M. Gordon, J. Appl. Phys. {\bf 87}, 8093 (2000).

\bibitem{rezek} Y. Rezek {\it et al.}, EPL {\bf 85}, 30008 (2009).

\bibitem{segal} D. Segal, Phys. Rev. Lett. {\bf 101}, 260601 (2008).

\bibitem{karen} A. E. Allahverdyan, K. Hovhannisyan and G. Mahler, Phys. Rev. E {\bf 81}, 051129 (2010).

\bibitem{olkin} A.W. Marshall and I. Olkin, {\it Inequalities: Theory
of Majorization and its Applications}, (Academic Press, New York,
1979).

\bibitem{q_t}J. Gemmer, M. Michel and G. Mahler, {\it Quantum Thermodynamics}
(Springer, NY, 2004).

\bibitem{landau}L.D. Landau and E.M. Lifshitz, {\it Statistical 
Physics, I}, (Pergamon Press Oxford, 1978). 

\bibitem{aa}A. E. Allahverdyan and Th. M. Nieuwenhuizen, Phys. Rev. E {\bf 71}, 046107 (2005).

\bibitem{abn}A.E. Allahverdyan, R. Balian and Th.M. Nieuwenhuizen, Europhys. Lett. {\bf 67}, 565 (2004).



\bibitem{lindblad}G. Lindblad, {\it Non-Equilibrium Entropy and
Irreversibility}, (D. Reidel, Dordrecht, 1983).

\bibitem{passivity} A. Lenard, J. Stat. Phys. {\bf 19}, 575 (1978).  

I.M. Bassett, Phys. Rev. A {\bf 18}, 2356 (1978).

W. Thirring, {\it Quantum Mechanics of Large Systems}, vol. 4 of {\it A Course in Mathematical Physics} (Springer-Verlag, Wien, 1980).

\bibitem{parrondo}J.M.R. Parrondo, C. Van den Broeck and R. Kawai, New J. Phys. {\bf 11}, 073008 (2009).

\bibitem{touchette} H. Touchette, R. S. Ellis, and B. Turkington, Physica A {\bf 335}, 518 (2004).

\bibitem{borgonovi} F. Borgonovi and F. M. Izrailev, Phys. Rev. E {\bf 62}, 6475 (2000).


\bibitem{kastner} M. Kastner and O. Schnetz, J. Stat. Phys. {\bf 122}, 1195 (2006).

\bibitem{leyvraz} A. Ramirez-Hernandez, H. Larralde and F. Leyvraz, Phys. Rev. E {\bf 78}, 061133 (2008).

\bibitem{campa}A. Campa, T. Dauxois and S. Ruffo, Phys. Rep. {\bf 480}, 57 (2009).

\bibitem{passo}
A. E. Allahverdyan and Th. M. Nieuwenhuizen, Physica A {\bf 305}, 542 (2002).

\bibitem{widom} J. Wu and A. Widom, Phys. Rev. E {\bf 57}, 5178 (1998).
\comment{ 
Standard statistical thermodynamic views of temperature fluctuations
predict a magnitude (??(?T)2?/T)??(kB/C) for a system with heat capacity
C. The extent to which low temperatures can be well defined is discussed
for those systems that obey the thermodynamic third law in the form
limop(T?0)C=0. Physical limits on the notion of very low temperatures
are exhibited for simple systems. Application of these concepts to bound
Bose condensed systems are explored, and the notion of bound boson
superfluidity is discussed in terms of the thermodynamic moment of
inertia. }

\bibitem{ja}T. Jahnke, S. Lanery, and G. Mahler, Phys. Rev. E {\bf 83}, 011109 (2011).

\comment{In this paper we present a quantum approach to the old problem
of temperature fluctuations. We start by observing that according to
quantum thermodynamics, fluctuations of intensive parameters like
temperature cannot exist. Furthermore, such parameters are not
observables, so their estimation has to be done indirectly. The
respective temperature estimate based on quantum measurements of the
energy is shown to fluctuate according to the well-known formula
?T2=kBT2/C, but only within a certain temperature range and if the
system is not too small. We also calculate the fourth-order correction
term, becoming important at higher temperatures. Finally we illustrate
our results with a concrete model of n spins.}

\bibitem{casimir1}
V. B. Bezerra, G. L. Klimchitskaya, V. M. Mostepanenko, and C. Romero, Phys. Rev. A {\bf 69}, 022119 (2004). 

\comment{Violation of the Nernst heat theorem in the theory of the thermal Casimir force between Drude metals

V. B. Bezerra, G. L. Klimchitskaya, V. M. Mostepanenko, and C. Romero, Phys. Rev. A 69, 022119 (2004) 

We give a rigorous analytical derivation of low-temperature behavior of
the Casimir entropy in the framework of the Lifshitz formula combined
with the Drude dielectric function. An earlier result that the Casimir
entropy at zero temperature is not equal to zero and depends on the
parameters of the system is confirmed, i.e., the third law of
thermodynamics (the Nernst heat theorem) is violated. We illustrate the
resolution of this thermodynamical puzzle in the context of the surface
impedance approach by several calculations of the thermal Casimir force
and entropy for both real metals and dielectrics. Different
representations for the impedances, which are equivalent for real
photons, are discussed. Finally, we argue in favor of the Leontovich
boundary condition which leads to results for the thermal Casimir force
that are consistent with thermodynamics. }


\bibitem{casimir2}S. A. Ellingsen, Phys. Rev. E {\bf 78}, 021120 (2008).

\comment{
Nernst's heat theorem for Casimir-Lifshitz free energy

Simen A. Ellingsen, Phys. Rev. E 78, 021120 (2008) 

Consideration of the Lifshitz expression for the Casimir free energy on
the real frequency axis rather than the imaginary Matsubara frequencies
as is customary sheds light on the ongoing debate regarding the
thermodynamical consistency of this theory in combination with common
permittivity models. It is argued that when permittivity is temperature
independent over a temperature interval including zero temperature, a
cavity made of causal material with continuous dispersion properties
separated by vacuum cannot violate Nernst's theorem (the third law of
thermodynamics). The purported violation of this theorem pertains to
divergencies in the double limit in which frequency and temperature
vanish simultaneously. While any model should abide by the laws of
thermodynamics within its range of applicability, we emphasize that the
Nernst heat theorem is a relevant criterion for choosing among candidate
theories only when these theories are fully applicable at zero
temperature and frequency. }

\bibitem{casimir3} M. Bordag and I. G. Pirozhenko, Phys. Rev. D {\bf 82}, 125016 (2010). 

\comment{Casimir entropy for a ball in front of a plane

M. Bordag and I. G. Pirozhenko, Phys. Rev. D 82, 125016 (2010) 

The violation of the third law of thermodynamics for metals described by
the Drude model and for dielectrics with finite dc conductivity is one
of the most interesting problems in the field of the Casimir effect. It
manifests itself as a nonvanishing of the entropy for vanishing
temperature. We review the relevant calculations for plane surfaces and
calculate the corresponding contributions for a ball in front of a
plane. In this geometry, these appear in much the same way as for
parallel planes. We conclude that the violation of the 3rd law is not
related to the infinite size of the planes. }

\bibitem{wald} R. M. Wald, Phys. Rev. D {\bf 56}, 6467 (1997).

G. Chirco {\it et al.}, Phys. Rev. D {\bf 82}, 104015 (2010).

\comment{

"Nernst theorem" and black hole thermodynamics

Robert M. Wald, Phys. Rev. D 56, 6467 (1997)

The Nernst formulation of the third law of ordinary thermodynamics
(often referred to as the "Nernst theorem") asserts that the entropy S
of a system must go to zero (or a "universal constant") as its
temperature T goes to zero. This assertion is commonly considered to be
a fundamental law of thermodynamics. As such, it seems to spoil the
otherwise perfect analogy between the ordinary laws of thermodynamics
and the laws of black hole mechanics, since rotating black holes in
general relativity do not satisfy the analogue of the "Nernst theorem."
The main purpose of this paper is to attempt to lay to rest the "Nernst
theorem" as a law of thermodynamics. We consider a boson (or fermion)
ideal gas with its total angular momentum J as an additional state
parameter, and we analyze the conditions on the single-particle density
of states, g(?,j), needed for the Nernst formulation of the third law to
hold. (Here, ? and j denote the single-particle energy and angular
momentum.) Although it is shown that the Nernst formulation of the third
law does indeed hold under a wide range of conditions, some simple
classes of examples of densities of states which violate the "Nernst
theorem" are given. In particular, at zero temperature, a boson (or
fermion) gas confined to a circular string (whose energy is proportional
to its length) not only violates the "Nernst theorem" also but
reproduces some other thermodynamic properties of an extremal rotating
black hole. }



\bibitem{leff} H. S. Leff, Phys. Rev. A {\bf 2}, 2368 (1970).

\comment{
Proof of the Third Law of Thermodynamics for Ising Ferromagnets

Harvey S. Leff, Phys. Rev. A 2, 2368 (1970)

The third law of thermodynamics is proved for a large class of Ising
models with generalized ferromagnetic many-body interactions. A
sufficient condition for the third law to hold is that the model have
nearest-neighbor couplings which are bounded from below by a positive
constant. The proof is based on a spin-correlation inequality of
Griffiths which implies a corresponding inequality for the bulk entropy
per spin. Ground-state degeneracy considerations are completely avoided. 
}

\bibitem{wu} Y. Chow and F. Y. Wu, Phys. Rev. B {\bf 36}, 285 (1987).

\comment{
Residual entropy and validity of the third law of thermodynamics in discrete spin systems

Yunshyong Chow and F. Y. Wu, Phys. Rev. B 36, 285 (1987) 

We study the validity of the third law of thermodynamics and the
occurrence of a nonzero residual entropy in discrete spin systems. For a
general classical spin system on a d-dimensional hypercubic lattice with
isotropic, translationally invariant, nearest-neighbor interactions, we
establish the following. (i) The necessary and sufficient condition for
the third law to hold. (ii) A lower bound on the residual entropy when
the third law is not valid. It is also established that the residual
entropy is nonzero for all d, if it is nonzero in any dimension.}

\bibitem{behn} U. Behn and V.A. Zagrebnov, J. Phys. A {\bf 21}, 2151 (1988).

\comment{One-dimensional
Markovian-field Ising model: physical properties
and characteristics of discrete stochastic mapping}

\bibitem{1d}
G. Watson, G. Canright and F. L. Somer, Phys. Rev. E {\bf 56}, 6459 (1997).

\comment{
Reasonable and robust Hamiltonians violating the third law of thermodynamics

Greg Watson, Geoff Canright, and Frank L. Somer, , Jr. Phys. Rev. E 56, 6459 (1997) 

It has recently been shown that the third law of thermodynamics is
violated by an entire class of classical Hamiltonians in one dimension,
over a finite volume of coupling-constant space, assuming only that
certain elementary symmetries are exact, and that the interactions are
finite ranged. However, until now, only the existence of such
Hamiltonians was known, while almost nothing was known of the nature of
the couplings. Here we show how to define the subvolume of these
Hamiltonians—a "wedge" W in a d-dimensional space—in terms of simple
properties of a directed graph. We then give a simple expression for a
specific Hamiltonian H* in this wedge, and show that H* is a physically
reasonable Hamiltonian, in the sense that its coupling constants lie
within an envelope that decreases smoothly, as a function of the range
l, to zero at l=r+1, where r is the range of the interaction. }

\bibitem{nisoli}C. Nisoli {\it et al.}, Phys. Rev. Lett. {\bf 98}, 217203 (2007). 

G.C. Lau {\it et al.}, Nature Physics {\bf 2}, 249 (2006).



\bibitem{glass}
M. Huang and J. P. Sethna, Phys. Rev. B {\bf 43}, 3245 (1991);
J. J. Brey and A. Prados, {\it ibid.} {\bf 43}, 8350 (1991); D. A.
Parshin and A. Wurger, {\it ibid.} {\bf 46}, 762 (1992).

\bibitem{glass1} S.M. Stishov, Sov. Phys. Usp. {\bf 31}, 52 (1988).


\bibitem{glass2} L.F. Cugliandolo {\it et al.}, Phys. Rev. E {\bf 55}, 3898 (1997). 

Th. M. Nieuwenhuizen, Phys. Rev. E {\bf 61}, 267 (2000) 

\bibitem{niki} A.F. Nikiforov and V.B. Uvarov, {\it Special Functions of Mathematical Physics} (Birkhauser, 1988).

\comment{Jill C Bonner and  Michael E Fisher,
The Entropy of an Antiferromagnet in a Magnetic Field
1962 Proc. Phys. Soc. 80 508 doi: 

The Ising model of an antiferromagnet exhibits an anomalous entropy peak
at a critical magnetic field which remains even at the absolute zero of
temperature, in violation of the third law of thermodynamics. By
investigating numerically the behaviour of finite chains of spins
interacting through the more general anisotropic Heisenberg coupling,
described by the Hamiltonian provided it is shown that the zero point
anomalous entropy is peculiar to the Ising model (\gamma = 0). If the
anisotropy is sufficiently great, however, an appreciable entropy peak
does persist down to low temperatures although in the anisotropic (pure
Heisenberg) limit the entropy has only a broad maximum which falls
steadily in height as T approaches zero.}


\comment{
Debye P 1926 Ann. Phys. 81 1154
Giauque W F 1927 J. Am. Chem. Soc. 49 1864 }

\bibitem{ve}
M. A. Nielsen and I. L. Chuang, {\it Quantum Computation and Quantum Information} (Cambridge University
Press, 2000).

V. Vedral, Rev. Mod. Phys. {\bf 74}, 197 (2002).


\bibitem{theo}Th.M. Nieuwenhuizen and A.E. Allahverdyan, Phys. Rev. E, {\bf 66}, 036102 (2002).

\end{thebibliography}
\end{document}